\documentclass[
twocolumn,
reprint,
superscriptaddress,
nofootinbib,
nobibnotes,
amsmath,
amssymb,
aps,
longbibliography,
prb,
showkeys
]{revtex4-2}
 
\usepackage[latin1]{inputenc}
\usepackage{amsmath}
\usepackage{amsfonts}
\usepackage{hyperref}
\usepackage{amssymb}
\usepackage{graphicx}
\usepackage{hyperref}
\usepackage{colortbl}
\usepackage{subfigure}
\usepackage{svg}
\usepackage{amsthm}
\usepackage{thmtools}
\usepackage[normalem]{ulem}
\usepackage{dsfont}
\usepackage{circuitikz}
\usepackage{cases}

\usepackage[T1]{fontenc}
\usepackage{mathptmx}
\usepackage{bbold}
\usepackage{scalerel}
\usepackage{xcolor}

\usepackage{mathtools}
\usepackage{dutchcal}

\declaretheorem[style=definition]{definition} 

\numberwithin{equation}{section}

\newcommand\numberthis{\addtocounter{equation}{1}\tag{\theequation}}

\newcommand{\hyref}[1]{\hyperref[#1]{\ref{#1}}}

\newcommand{\orange}[1]

\renewcommand{\thesection}{\arabic{section}}
\renewcommand{\thesubsection}{.\arabic{subsection}}

\newcommand{\thenewsubsection}{\thesection.\arabic{subsection}}
\newcommand{\thenewsubsubsection}{\thesection\thesubsection.\arabic{subsubsection}}

\makeatletter
\def\@hangfrom@section#1#2#3{\@hangfrom{#1#2}#3}
\def\@hangfroms@section#1#2{#1#2}
\makeatother

\usepackage{titlesec}

\titleformat{\section}  
  {\fontsize{12}{14}\bfseries} 
  {\thesection} 
  {1em} 
  {\centering} 
  [] 

\titleformat{\subsection}  
  {\fontsize{10}{12}\bfseries} 
  {\thenewsubsection} 
  {2em} 
  {\centering} 
  [] 

\titleformat{\subsubsection}  
  {\fontsize{10}{12}\itshape} 
  {\thenewsubsubsection} 
  {2em} 
  {\centering} 
  [] 
\begin{document}

\title{Echo State and Band-pass Networks with aqueous memristors: leaky reservoir computing with a leaky substrate}

\author{T. M. Kamsma}
\affiliation{Mathematical Institute, Utrecht University, Budapestlaan 6, 3584 CD Utrecht, The Netherlands}
\affiliation{Institute for Theoretical Physics, Utrecht University,  Princetonplein 5, 3584 CC Utrecht, The Netherlands}
\author{J. J. Teijema}%
\affiliation{Department of Methodology and Statistics, Utrecht University, Padualaan 14, 3584 CH Utrecht, The Netherlands}
\author{R. van Roij}
\affiliation{Institute for Theoretical Physics, Utrecht University, Princetonplein 5, 3584 CC Utrecht, The Netherlands}
\author{C. Spitoni}%
\affiliation{Mathematical Institute, Utrecht University, Budapestlaan 6, 3584 CD Utrecht, The Netherlands}

\date{\today}

\begin{abstract}
Recurrent Neural Networks (RNN) are extensively employed for processing sequential data such as time series. Reservoir computing (RC) has drawn attention as an RNN framework due to its fixed network that does not require training, making it an attractive platform for hardware based machine learning. We establish an explicit correspondence between the well-established mathematical RC implementations of Echo State Networks and Band-pass Networks with Leaky Integrator nodes on the one hand and a physical circuit containing iontronic simple volatile memristors on the other. These aqueous iontronic devices employ ion transport through water as signal carriers, and feature a voltage-dependent (memory) conductance. The activation function and the dynamics of the Leaky Integrator nodes naturally materialise as the (dynamic) conductance properties of iontronic memristors, while a simple fixed local current-to-voltage update rule at the memristor terminals facilitates the relevant matrix coupling between nodes. We process various time series, including pressure data from simulated airways during breathing that can be directly fed into the network due to the intrinsic responsiveness of iontronic devices to applied pressures. We accomplish this by employing established physical equations of motion of iontronic memristors for the internal dynamics of the circuit.
\end{abstract}

\maketitle

\textbf{Reservoir computing (RC) is a proven method for processing temporal data and has drawn more recent attention as a suitable framework for hardware based machine learning. Echo State and Band-pass Networks are extensively studied implementations of RC. We propose a novel physical circuit design, based on fluidic iontronic memristors, that provides a one-to-one correspondence with the mathematical descriptions of these RC paradigms. Using the underlying equations of motion of these fluidic devices, we process several time series, including simulated respiratory pressure waveforms, exploiting iontronics' intrinsic sensitivity to applied pressures. Our direct physical (iontronic) realization of these established RC implementations offers a blueprint for physically embedded temporal processing with an emerging substrate.}

\section{\label{sec:introduction}Introduction}
Reservoir computing (RC) has gained significant attention as a Recurrent Neural Network paradigm for processing temporal data \cite{Nakajima2021NaturalApplications}. RC employs a fixed high-dimensional reservoir (i.e.\ a dynamical system with many internal states) whose dynamics are driven by input signals, with the benefit that only a simple readout function requires training for classification tasks. Although the rise of computational capacity for training in the past few years has somewhat mitigated this benefit, new attention has recently been drawn to RC for hardware-based implementations as the fixed nature of the reservoir circumvents complicated internal tuning of the RC circuit \cite{Cucchi2022Hands-onImplementation}. Recent research, for instance, has explored the use of physical systems such as electronic, electrochemical, optical, and mechanical devices \cite{Liang2024PhysicalElectronics,Cucchi2022Hands-onImplementation}. However, although the use of physical substrates is informed by the established mathematical frameworks for software RC, establishing a deeper physical equivalence is challenging.

In this work, we establish an explicit one-to-one correspondence between the physical equations of iontronic memristors placed within a peripheral circuit on the one hand and the governing equations of Echo State Networks (ESNs) and Band-pass Networks (BPNs) with Leaky Integrator nodes (LI-ESNs and LI-BPNs, respectively) on the other. Iontronics exploit aqueous ionic and molecular transport, akin to the brain's medium, and can therefore provide striking similarities with the brain in neuromorphic computing implementations \cite{Noy2023FluidDevices,Noy2023NanofluidicSplash,Hou2023LearningNanofluidics,Yu2023BioinspiredComputing,Xie2022PerspectiveApplication,Fan2025EmergingComputation,SuwenLaw2025RecentComputing, Lv2025AdvancementsDesign}, including RC \cite{Kamsma2023Brain-inspiredNanochannels,Portillo2025NeuromorphicDiodes}. Additionally, the easily tunable memory timescales of iontronic platforms\cite{Cervera2024ModelingDiodes,Kamsma2023Brain-inspiredNanochannels,Kamsma2023IontronicMemristors,Zhang2024GeometricallySystems,Kamsma2024AdvancedCircuit} naturally match the relatively slow timescales found in natural or biological signals, something that is challenging within conventional fast solid-state devices\cite{Chicca2020ASystems}. Moreover, we show how the pressure-dependence of these fluidic systems enables direct conversion of a biological pressure signal to circuit input without any intervention or interaction required from outside the network.

In this work we \textit{(i)} propose a circuit based on the emerging ``leaky'' substrate of aqueous iontronics that would be capable of advanced RC applications, \textit{(ii)} propose a one-to-one correspondence between this physical (iontronic) circuit and the well-established mathematical LI-ESN and LI-BPN descriptions, \textit{(iii)} leverage the unique property of slow easily tunable memory timescales of iontronic memristors, and \textit{(iv)} exploit the intrinsic pressure responsiveness of iontronics to directly convert pressure signal inputs on-chip. Due to the equivalence between our proposed physical device and the abstract mathematical framework, the (extensive) theoretical results previously derived for ESNs and BPNs \cite{Jaeger2002AdaptiveNetworks,Jaeger2007OptimizationNeurons,Jaeger2010The1,Yildiz2012Re-visitingProperty,Lukosevicius2012ANetworks,Whiteaker2021LeakySets,Lin2024AComputing,Sun2024AApplication,Wyffels2008Band-passComputing} can be directly translated to the proposed physical circuit, without requiring us to reinvent the wheel for RC on a physical substrate. All code used for our results is available online at \cite{pyontronics}. Our results not only advance the theoretical understanding of RC in physical systems but also provide a pathway for the development of new (iontronic) hardware-based RC implementations.

\begin{figure*}[ht!]
\centering
     \includegraphics[width=1\textwidth]{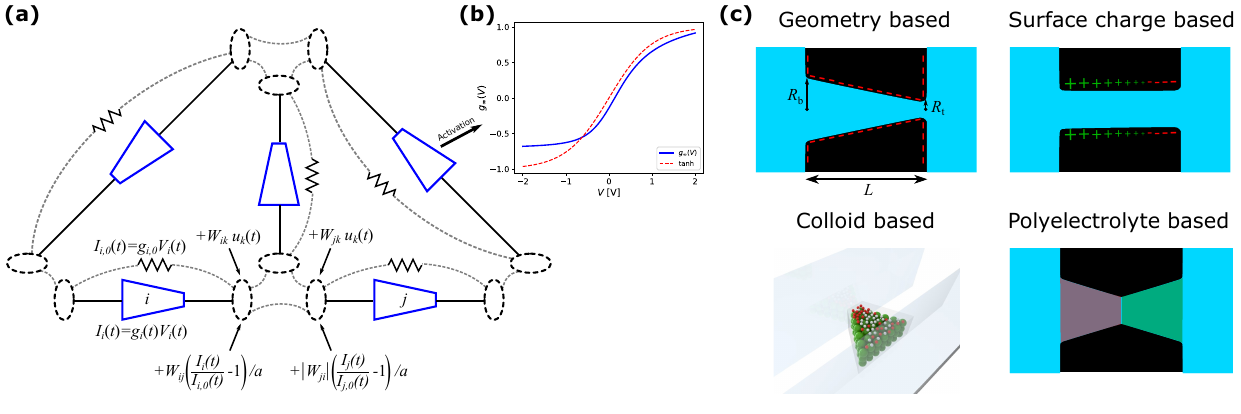}
        \caption{\textbf{(a)} Schematic of a physical Leaky Integrator Echo State or Band-pass network circuit containing (iontronic) Simple Volatile Memristors. Resistors are connected in parallel to the memristors with conductances equal to the equilibrium conductance of their respective parallel memristors. The terminals at either end of the memristors convert the incoming currents of neighbouring memristor-resistor pairs to voltages of the next time step. For simplicity, only the inputs to and coupling between $i$ and $j$ memristors are included. The connections in this schematic are for illustrative purposes only and do not represent a specific circuit topology used in this work. \textbf{(b)} Dimensionless conductance $\mathcal{g}$ (blue), acting as a physical activation function in the circuit, compared to the standard ESN $\text{tanh}$ activation function \cite{Jaeger2007OptimizationNeurons} (red). \textbf{(c)} Several candidate iontronic Simple Volatile Memristors where the conductance memory is geometry-based (top left) \cite{Kamsma2023IontronicMemristors,Cervera2024ModelingDiodes}, surface charge-based (top-right) \cite{Kamsma2023UnveilingIontronics,Kamsma2024AdvancedCircuit}, colloid-based \cite{Kamsma2023Brain-inspiredNanochannels} (bottom left), and polyelectrolyte-based (bottom right) \cite{Zhang2024GeometricallySystems}. All results in this work are from geometry-based iontronic memristors \cite{Kamsma2023IontronicMemristors}.}
        \label{fig:Fig1}
\end{figure*}
\section{Echo State and Band-pass Networks}
We consider LI-ESNs that converts a $K$-dimensional input ${\bf u}(t)$ with $N$ reservoir neurons at state ${\bf x}(t)$ to an $L$-dimensional output ${\bf y}(t)$ at time $t$, here without output feedback and without direct input-to-output coupling. Such an LI-ESN is governed by \cite{Jaeger2007OptimizationNeurons}
\begin{align}
    \dot{\textbf{x}}=&\frac{1}{c}\left(-a\textbf{x}+f\big(\textbf{W}^{\text{in}}\textbf{u}+\textbf{W}\textbf{x}\big)\right),\label{eq:ESN_LIN}\\ \textbf{y}=&\textbf{W}^{\text{out}}\textbf{x}.\label{eq:ESN_out}
\end{align}
Here $c\in\mathds{R}$ is a global relaxation time parameter, $a\in\mathds{R}$ is the leaking rate, $\textbf{u}=\textbf{u}(t)\in\mathds{R}^{K}$ is the input, $\textbf{x}=\textbf{x}(t)\in\mathds{R}^{N}$ the state of the reservoir neurons, $f:\mathds{R}\to\mathds{R}$ is a sigmoidal activation function that is applied element-wise to its input, and  $\textbf{y}=\textbf{y}(t)\in\mathds{R}^{L}$ is the output vector. Here we will set $f(x)=\tanh(x)$, a standard choice\cite{Jaeger2007OptimizationNeurons}, and $g$ will be the identity function such that $\textbf{y}=\textbf{W}^{\text{out}}\textbf{x}$. The input and reservoir states are coupled through the input matrix $\textbf{W}^{\text{in}}\in\mathds{R}^{N\times K}$, the internal matrix $\textbf{W}\in\mathds{R}^{N\times N}$, and the output matrix $\textbf{W}^{\text{out}}\in\mathds{R}^{L\times N}$. Notably, only $\textbf{W}^{\text{out}}$ needs to be found through training, which we here perform via ridge regression\cite{Hoerl1970RidgeProblems}, with the other matrices being initialised randomly. 

In its most general form there is also an output-feedback term $\textbf{W}^{\text{fb}}\mathbf{y}$ in the argument of $f$, which is not of relevance for the input processing we focus on here \cite{Jaeger2007OptimizationNeurons}. Additionally, the output can be transformed by a function $g:\mathds{R}\to\mathds{R}$ that is applied element-wise to $\textbf{W}^{\text{out}}\textbf{x}$, such that $\textbf{y}=g\left(\textbf{W}^{\text{out}}\textbf{x}\right)$. Lastly, in general one can also directly couple the input to the output according to $\textbf{y}=g\left(\textbf{W}^{\text{out}}[\textbf{x};\textbf{u}]\right)$ with $[;]$ denoting vector concatenation (in this case $\textbf{W}^{\text{out}}\in\mathds{R}^{L\times (K+N)}$). These last two generalisations are omitted here to simplify the physical circuit realization.

The physical circuit we will introduce in Sec.~\ref{sec:ESNcircuit} is in principle described by continuous equations as Eq.~(\ref{eq:ESN_LIN}). However, inputs $\mathbf{u}$ are often discrete-time sampled. Additionally, here we simulate the circuit using physical equations of the internally used iontronic memristors, but this will also require discretization of the underlying equations. Using Euler discretization with (time) stepsize $\delta$ and $t=n\delta$ with $n\in\mathds{N}$, we see that Eq.~(\ref{eq:ESN_LIN}) for $n\geq 0$ becomes\cite{Jaeger2007OptimizationNeurons}
\begin{align*}
    \textbf{x}(n+1)=&\left(1-\frac{a\delta}{c}\right)\textbf{x}(n)\label{eq:ESN_LIN_discrete}\numberthis\\
    &+\frac{\delta}{c}f\left(\textbf{W}^{\text{in}}\textbf{u}(n\delta)+\textbf{W}\textbf{x}(n)\right).
\end{align*}
\subsection{Echo State property}
A key stability property of ESNs is the \textit{echo state property} (ESP), which is defined by\cite{Jaeger2007OptimizationNeurons}
\begin{definition}\label{def:ESP}
An ESN with reservoir states $\mathbf{x}(n)$ has the echo state property if for any compact $C\subset \mathds{R}^{K}$ and any two starting states $\mathbf{x}(0)$ and $\mathbf{x}^{\prime}(0)$, there exists a sequence $(\delta_h)_{h=0,1,2,...}$ that converges to 0 such that for any input sequence $(\mathbf{u}(n))_{n=0,1,2,...}\subseteq C$ it holds that $\|\mathbf{x}(h)-\mathbf{x}^{\prime}(h)\|\leq\delta_{h}$.
\end{definition}
Heuristically, Def.~\ref{def:ESP} tells us that an ESN with the ESP ``forgets'' its initial state at a rate independent from the input sequence or the precise initial state.

There are various constraints that guarantee the echo state property in Leaky Integrator ESNs \cite{Jaeger2007OptimizationNeurons,Yildiz2012Re-visitingProperty}. The condition we will use here is that the spectral radius $\rho(\mathbf{M})$ of $\mathbf{M}=(\delta/c)\left|\mathbf{W}\right|+(1-a\delta/c)\mathbf{I}$ satisfies $\rho(\mathbf{M})<1$. A simple algorithm for constructing an internal weight matrix that guarantees the echo state property is given by \cite{Yildiz2012Re-visitingProperty}
\begin{enumerate}
    \item Generate a random matrix $\mathbf{W}$ with only non-negative elements $w_{ij}\geq 0$.
    \item Rescale $\mathbf{W}$ such that the spectral radius $\rho\left(\mathbf{M}\right)$ of the matrix $\mathbf{M}=(\delta/c)\mathbf{W}+(1-a\delta/c)\mathbf{I}$ satisfies $\rho\left(\mathbf{M}\right)<1$.
    \item Change the sign of a desired number of elements $w_{ij}$.
\end{enumerate}
Internal weight matrices $\mathbf{W}$ with spectral radii smaller than 1 often also display the ESP, but this is not a guarantee \cite{Yildiz2012Re-visitingProperty}. Alternatively, one could generate a random $\mathbf{W}$ and then check if $\rho(\mathbf{M})<1$ afterwards. The available code\cite{pyontronics} provides aforementioned algorithm as an optional setting to guarantee the ESP. Lastly, $a\frac{\delta}{c}\leq 1$ is a natural constraint.

\subsection{Band-pass network}
A LI-BPN is similar to a LI-ESN as in Eqs.~(\ref{eq:ESN_LIN}) and (\ref{eq:ESN_out}), with one key difference. Each node can individually be designed to be sensitive to certain frequencies, which we implement here by providing each node with its own characteristic relaxation timescale $c_i$ \cite{Wyffels2008Band-passComputing,Holzmann2010EchoReadout}, as opposed to a single global timescale $c$ for the entire network. So the parameter $c\in\mathds{R}$ is replaced by a vector $\mathbf{c}\in\mathds{R}^{N}$ and Eq.~(\ref{eq:ESN_LIN}) naturally becomes
\begin{align}
    \dot{\textbf{x}}=&\left(-a\textbf{x}+f\left(\textbf{W}^{\text{in}}\textbf{u}+\textbf{W}\textbf{x}\right)\right)\oslash\mathbf{c},\label{eq:BP_LIN}
\end{align}
with $\oslash\mathbf{c}$ element-wise (Hadamard) division.

\section{Physical LI-ESN with iontronic memristors}
Consider the circuit schematically drawn in Fig.~\ref{fig:Fig1}(a) containing (iontronic) memristors (blue), which we will describe in more detail in Sec.~\ref{sec:SVM}. The voltages at the terminals (dashed ellipses) obey a fixed local current-to-voltage update rule, which we will describe in detail in Sec.~\ref{sec:currenttovoltage}. We will show in Sec.~\ref{sec:ESNequiv} that the physical circuit design in Fig.~\ref{fig:Fig1}(a) is equivalent to the general mathematical LI-ESN description as in Eq.~(\ref{eq:ESN_LIN}). Lastly, in Sec.~\ref{sec:band-pass} we will extend this equivalence to LI-BPNs as in Eq.~(\ref{eq:BP_LIN}).
\subsection{Physical ESN circuit}\label{sec:ESNcircuit}

\subsubsection{(Iontronic) Simple Volatile Memristors}\label{sec:SVM}
Memristors, characterised by their history-dependent conductance, have drawn major interest as fundamental devices for neuromorphic computing architectures \cite{Schuman2022OpportunitiesApplications}. Consequently, many different types of memristors with various conductance memory features exist \cite{Sangwan2020NeuromorphicMaterials,Zhu2020ADevices,Schuman2017AHardware}. Inspired by the brain's aqueous medium and ionic signal carriers, iontronics that rely on ionic transport in an aqueous environment are emerging as a substrate for neuromorphic computing implementations  \cite{Noy2023FluidDevices,Noy2023NanofluidicSplash,Hou2023LearningNanofluidics,Yu2023BioinspiredComputing,Xie2022PerspectiveApplication}. Of importance to this work is that various iontronic devices also feature a coupling between their electric properties and applied pressures \cite{Jubin2018DramaticNanopores,Boon2022Pressure-sensitiveGeometry,Barnaveli2024Pressure-GatedProcessing}, where e.g.\ an applied pressure can drive a so-called electric \textit{streaming current} \cite{VanDerHeyden2005StreamingChannel,Werkhoven2020CoupledSystems}.

The (iontronic) memristors schematically drawn in blue in Fig.~\ref{fig:Fig1}(a) are \textit{Simple Volatile Memristors} (SVMs) \cite{Kamsma2024ACircuits}. It has been demonstrated theoretically \cite{Kamsma2023IontronicMemristors,Kamsma2023UnveilingIontronics} and experimentally \cite{Kamsma2023Brain-inspiredNanochannels,Cervera2024ModelingDiodes} that various fluidic iontronic memristors behave as SVMs \cite{Kamsma2024ACircuits}, of which the electric conductance $g_i(t)$ is time-dependent and obeys the equation of motion (EOM)
\begin{align}\label{eq:condEOM}
    \dfrac{\text{d}g_i}{\text{d}t}=\frac{g_{i,\infty}(V_{i}(t))-g_{i}(t)}{\tau_i}.
\end{align}
Here $g_{i,\infty}(V_{i})$ is the steady-state conductance for a given voltage $V_i$, which is typically a sigmoidal function around the equilibrium conductance $g_{i,0}=g_{i,\infty}(0)$. It has been theoretically derived \cite{Kamsma2023IontronicMemristors,Kamsma2023UnveilingIontronics} and experimentally observed \cite{Kamsma2023Brain-inspiredNanochannels,Cervera2024ModelingDiodes,Zhang2024GeometricallySystems} that the intrinsic memory timescale $\tau_i$ of various iontronic SVMs scales quadratically with the device length $L_i$ according to
\begin{align}
    \tau_i\propto \frac{L_i^2}{D},\label{eq:ts}
\end{align}
where $D$ is the ionic diffusion coefficient, assumed equal for all ionic species of the aqueous electrolyte involved. Due to its dependence on $L_i$, the timescale $\tau_i$ can be individually chosen for each SVM across a wide range, which we will use in Sec.~\ref{sec:band-pass} to implement the individual relaxation times in LI-BPNs. A variety of different iontronic SVMs are candidates for the circuit we propose here, including channels where the conductance memory is geometry-based \cite{Kamsma2023IontronicMemristors,Cervera2024ModelingDiodes}, surface charge-based \cite{Kamsma2023UnveilingIontronics}, colloid-based \cite{Kamsma2023Brain-inspiredNanochannels}, and polyelectrolyte-based \cite{Zhang2024GeometricallySystems}, as schematically depicted in Fig.~\ref{fig:Fig1}(c). Additionally, the power consumption of iontronic memristive devices can be extremely small, as low as order $10$ fW per channel \cite{Shi2023UltralowMemristor} (assuming order $~1$ V driving force).

For our network demonstrations here we chose to consider conical channel SVMs \cite{Kamsma2023IontronicMemristors}, but the results are representative of any SVM with a sigmoidal steady-state conductance. Specifically, we consider microfluidic channels as drawn in the top left of Fig.~\ref{fig:Fig1}(c) with a base radius $R_{\mathrm{b}}=200~\text{nm}$, a tip radius $R_{\mathrm{t}}=50~\text{nm}$, a charge on the channel's surface of $-2.4\cdot10^{-22}~C/\text{nm}^2$, filled with an aqueous 1:1 electrolyte with equilibrium ion concentrations of 0.1 mM for both the positive and negative ions. The conductance of the channel is voltage-dependent and shown in Fig.~\ref{fig:Fig1}(b) in blue (normalised and centered around 0). The proportionality constant in Eq.~(\ref{eq:ts}) can vary between iontronic devices, but the channels \cite{Kamsma2023IontronicMemristors} feature the relation $\tau_i=\frac{L_i^2}{12D}$, with lengths that can be fabricated from nm lengths \cite{Storm2003FabricationPrecision,Yang2016PolarizationRectification} all the way to mm length\cite{Zhou2023NanofluidicNanochannels} scales, theoretically corresponding to a broad timescale range from $\sim10^{-9}~\text{s}$ up to $\sim10^{3}~\text{s}$ domains. While experimental evidence for the full range is still limited, the order $\sim0.1-1~\text{s}$ timescales that we will use here have been observed experimentally \cite{Wang2012TransmembraneTransport,Kamsma2023Brain-inspiredNanochannels,Cervera2024ModelingDiodes}. The lengths $L_i$ of the channels vary between different network applications, and even within individual networks for BPNs, to implement the different timescales $\tau_i$ as per Eq.~(\ref{eq:ts}). The full detailed physics and remaining parameters are described in Appendix \ref{sec:appendix_cone}.

\subsubsection{Current-to-voltage update rule}\label{sec:currenttovoltage}
Consider the circuit schematically drawn in Fig.~\ref{fig:Fig1}(a), where voltage terminals (dashes ellipses) connect parallel pairs of an Ohmic resistor and an SVM, here in the form of cone-shaped iontronic microfluidic channels \cite{Kamsma2023IontronicMemristors} (blue). Memristors are two-terminal devices with $V_{i,\text{t}}$ and $V_{i,\text{b}}$ the voltages at the tip and base terminal, respectively, defined such that the voltage $V_{i}=V_{i,\text{t}}-V_{i,\text{b}}$ over the SVM increases the conductance for positive $V_{i}$. Between the terminal pairs two currents flow in parallel, a current $I_i=g_i(t)V_i(t)$ through the SVM and a current $I_{i,0}=g_{i,0}V_i(t)$ through the resistor with a fixed conductance $g_{i,0}$. The terminals obey the same update rule for the tip and base voltages $V_{i,t}$ and $V_{i,b}$, which depends on the currents $I_j$ and $I_{j,0}$ of the neighboring terminals according to
\begin{align}
    V_{i,\text{t}}=&\sum_{j: W_{ij}>0} W_{ij}\left(\frac{I_{j}}{I_{j,0}}-1\right)a^{-1}+\sum_{j: W^{\text{in}}_{ij}>0} W^{\text{in}}_{ij}u_{j}(t),\label{eq:VoltRule}\\
   V_{i,\text{b}}=&\sum_{j: W_{ij}<0} \left|W_{ij}\right|\left(\frac{I_{j}}{I_{j,0}}-1\right)a^{-1}+\sum_{j: W^{\text{in}}_{ij}<0}\left|W^{\text{in}}_{ij}\right|u_{j}(t).
\end{align}
Here $W_{ij}$, $W^{\text{in}}_{ij}$, and $a$ are fixed and known \textit{a priori}, $I_{j}(t)$ and $I_{0,j}(t)$ are physical currents that only need to be measured locally. Lastly, $u_{j}(t)$ is the dynamic input. Depending on the input type, the input can feature an additional scaling factor $s^{\text{in}}\in\mathds{R}$ such that $u_{j}(t)=s^{\text{in}}\tilde{u}_{j}(t)$. This factor can fix the units and ensure the input stays within a reasonable $\sim\pm1~V$ voltage regime. Notably, in Sec.~\ref{sec:pressure} $u_j(t)$ will receive its own local (pressure-to-)current-to-voltage update rule, where we analyze biological pressure signals as inputs by placing additional microfluidic channels between the pressure source and the SVM terminals. In such microfluidic channels, pressures are known to drive (electrical) \textit{streaming currents} through the channels \cite{VanDerHeyden2005StreamingChannel,Werkhoven2020CoupledSystems}, which can then be converted according to a current-to-voltage update rule similar to Eq.(\ref{eq:VoltRule}), thereby providing a direct physical conversion between a biological signal and the ESN or BPN input without any intervention or interaction required from outside the network. 

Although some functionality is assumed for the peripheral circuits at the terminals, this concerns only a straightforward conversion of locally measured currents to voltages with some a priori known fixed parameters. The use of peripheral circuitry for current-to-voltage conversions is relatively standard within neuromorphics, e.g.\ in the common neuromorphic circuits of coupled crossbar arrays that emulate artificial neural network current-to-voltage converters are employed to transform one array's current outputs to another array's voltage inputs \cite{Li2018EfficientNetworks}.

\subsection{Physical circuit and LI-ESN equivalence}\label{sec:ESNequiv}
For simplicity, let us initially consider all SVMs have equal length $L_i=L$ and therefore equal timescales $\tau_i=\tau$. In Sec.~\ref{sec:band-pass} we will make the straightforward extension to a range of timescales $\pmb{\tau}\in\mathds{R}^N$, equivalent to a Band-pass Network. We will show that the physical circuit design in Fig.~\ref{fig:Fig1}(a) is equivalent to the general mathematical LI-ESN description as in Eq.~(\ref{eq:ESN_LIN}). The circuit will thereby be endowed with all its relevant derived properties, capabilities, and understandings \cite{Jaeger2002AdaptiveNetworks,Jaeger2007OptimizationNeurons,Jaeger2010The1,Yildiz2012Re-visitingProperty,Lukosevicius2012ANetworks,Whiteaker2021LeakySets,Lin2024AComputing,Sun2024AApplication} , while the actual dynamics emerge from the intrinsic physics of the circuit, rather than numerically solving Eq.~(\ref{eq:ESN_LIN}) in software. Specifically, the conductance EOM Eq.~(\ref{eq:condEOM}) and steady state conductance $g_{i,\infty}(V_i)$ naturally assume the role of the ESN dynamics and the activation function. Furthermore, the relative polarity of the voltage depending on the orientation of the SVMs provides a natural method to encode either positive or negative (adjacency) weight elements.

 We introduce a straightforward conversion to a dimensionless conductance $\mathcal{g}_{i}(t)$ normalized by $g_{i,0}=g_{i,\infty}(0)$ the equilibrium conductance
\begin{align}
    \mathcal{g}_{i}(t)=&\frac{g_{i}(t)-g_{i,0}}{ag_{i,0}},\\
    \mathcal{g}_{i,\infty}(V_i)=&\frac{g_{i,\infty}(V_i)-g_{i,0}}{g_{i,0}}\approx \tanh(V_i).
\end{align}
We stress that all results presented in this work exclusively use the physical function $g_{i,\infty}(V_i)$ \cite{Kamsma2023IontronicMemristors} for the activation function $\mathcal{g}_{i,\infty}(V_i)$, shown in blue in Fig.~\ref{fig:Fig1}(b), alongside the function $\tanh(V_i)$ (red). The abovementioned similarity $\mathcal{g}_{i,\infty}(V_i)\approx \tanh(V_i)$ only serves to support the equivalence to LI-ESNs.

The EOM of $\mathcal{g}_i(t)$ is straightforwardly found through Eq.~(\ref{eq:condEOM}) as follows
\begin{align*}
    \dfrac{\text{d}\mathcal{g}_i}{\text{d}t}=&\frac{1}{ag_{i,0}}\frac{\left(g_{i,\infty}(V_{i})-g_{i,0}\right)-\left(g_{i}(t)-g_{i,0}\right)}{\tau}\\
    =&\frac{\mathcal{g}_{i,\infty}(V_{i})-a\mathcal{g}_{i}(t)}{a\tau}.
\end{align*}
In vector-notation this becomes
\begin{align}\label{eq:dimlessgeom}
    \dfrac{\text{d}\mathbcal{g}}{\text{d}t}=\frac{\mathcal{g}_{\infty}(\textbf{V})-a\mathbcal{g}(t)}{a\tau},
\end{align}
where $\mathcal{g}_{\infty}(\textbf{V})$ is applied element-wise to $\textbf{V}$.

Memristors are two terminal devices with voltages $V_{i,\text{t}}$ and $V_{i,\text{b}}$ at either terminal, respectively, here defined such that the voltage $V_{i}=V_{i,\text{t}}-V_{i,\text{b}}$ over the SVM increases the conductance for positive voltages. Both terminals have identical voltage update rules, which we now show are coupled to $\mathcal{g}$ as follows
\begin{align*}
    V_{i,\text{t}}=&\sum_{j: W_{ij}>0} W_{ij}\left(\frac{I_{j}}{I_{j,0}}-1\right)a^{-1}+\sum_{j: W^{\text{in}}_{ij}>0} W^{\text{in}}_{ij}u_{j}(t)\\
    =&\sum_{j: W_{ij}>0} W_{ij}\frac{g_{j}(t)V_{j}-g_{j,0}V_{j}}{ag_{j,0}V_{j}}+\sum_{j: W^{\text{in}}_{ij}>0} W^{\text{in}}_{ij}u_{j}(t)\\
    =&\sum_{j: W_{ij}>0} W_{ij}\mathcal{g}_{j}+\sum_{j: W^{\text{in}}_{ij}>0} W^{\text{in}}_{ij}u_{j}.
\end{align*}
Similarly
\begin{align*}
    V_{i,\text{b}}=&\sum_{j: W_{ij}<0} \left|W_{ij}\right|\mathcal{g}_{j}(t)+\sum_{j: W^{\text{in}}_{ij}<0}\left|W^{\text{in}}_{ij}\right|u_{j}(t).
\end{align*}

Therefore, the voltage over the SVM is given by
\begin{align*}
    V_{i}=&V_{i,\text{t}}-V_{i,\text{b}}\numberthis\label{eq:SVMvolt}\\
    =&\sum_{j: W_{ij}>0} W_{ij}\mathcal{g}_{j}-\sum_{j: W_{ij}<0} \left|W_{ij}\right|\mathcal{g}_{j}\\
    &+\sum_{j: W^{\text{in}}_{ij}>0} W^{\text{in}}_{ij}u_{j}-\sum_{j: W^{\text{in}}_{ij}<0} \left|W^{\text{in}}_{ij}\right|u_{j},
\end{align*}
where we now see that negative weights are naturally encoded through the voltage sign reversal. Moreover, we see that the current-to-voltage conversion rule in Eq.~(\ref{eq:VoltRule}) is equivalent to a matrix multiplication with the dimensionless conductances $\mathcal{g}_i(t)$.

Compactly, Eq.~(\ref{eq:SVMvolt}) can be written in matrix vector notation as
\begin{align}\label{eq:SVMvoltVec}
\textbf{V}=\textbf{W}\mathbcal{g}+\textbf{W}^{\text{in}}\textbf{u},
\end{align}
such that $\mathbcal{g}$ evolves according to
\begin{align}\label{eq:EOM_SVM_LIN}
    \dfrac{\text{d}\mathbcal{g}}{\text{d}t}=\frac{\mathcal{g}_{\infty}(\textbf{W}^{\text{in}}\textbf{u}+\textbf{W}\mathbcal{g})-a\mathbcal{g}(t)}{a\tau}
\end{align}
which we recognize as (the arguments inside) the activation function $f$ in Eq.~(\ref{eq:ESN_LIN}). Therefore, the straightforward current-to-voltage update role described in Eq.~(\ref{eq:VoltRule}) facilitates the matrix coupling between the nodes. This does require that the voltages can be adjusted quasi-instantaneously compared to the timescale $\tau$ of the SVM. To complete the equivalence to Eq.~(\ref{eq:ESN_LIN}), let us consider the identifications
\begin{align*}
    \mathcal{g}_{\infty}(x)\approx \tanh(x) \leftrightarrow& f(x)\\
    \mathbcal{g}(t) \leftrightarrow& \mathbf{x}(t)\\
    a\tau\leftrightarrow&c.
\end{align*}
We now see that Eq.~(\ref{eq:EOM_SVM_LIN}) is identical to Eq.~(\ref{eq:ESN_LIN}), while being completely physically facilitated in the circuit shown in Fig.~\ref{fig:Fig1}(a).

Above we described how the dynamics of the circuit depicted in Fig.~\ref{fig:Fig1}(a) are equivalent to the ESN dynamics as per Eq.~(\ref{eq:ESN_LIN}). Moreover, since applying the input and reading the output are performed by standard matrix multiplications $\textbf{W}^{\text{in}}\textbf{u}$ and $\textbf{W}^{\text{out}}\textbf{x}$ respectively, these actions too can be physically realised using crossbar arrays, which too could be implemented using ionic devices \cite{Liu2025Resistance-RestorableChip,VanDeBurgt2018OrganicComputing,Kazemzadeh2025AllArray,Xu2025Angstrom-Scale-ChannelComputing,Zhang2022AdaptiveConductor,Hu2023AnComputing,VanDeBurgt2017AComputing}. Therefore, excitingly, the full (ionic) hardware implementation of our LI-ESN circuit should be directly physically possible.

\subsection{Physical LI-BPN}\label{sec:band-pass}
Because each SVM can straightforwardly be designed to feature its own timescale $\tau_i\propto L_i^2/D$ by varying the length $L_i$ of the individual devices, we can easily go beyond ESNs to BPNs, which are known to perform considerably better on input tasks that feature components that span multiple frequencies \cite{Wyffels2008Band-passComputing}.

Varying the lengths between the different SVMs corresponds to converting $\tau$ to a vector $\tau\rightarrow\pmb{\tau}\in\mathds{R}^N$ such that Eq.~(\ref{eq:EOM_SVM_LIN}) becomes  
\begin{align}\label{eq:EOM_SVM_BandPass}
    \dfrac{\text{d}\mathbcal{g}}{\text{d}t}=\frac{\mathcal{g}_{\infty}(\textbf{W}^{\text{in}}\textbf{u}+\textbf{W}\mathbcal{g})-a\mathbcal{g}(t)}{a}\oslash\pmb{\tau},
\end{align}
with $\oslash\pmb{\tau}$ element-wise (Hadamard) division. We note that the dimensionless $\mathcal{g}_{i,\infty}(V)$ is independent from $L_i$, so the extension from Eq.~(\ref{eq:EOM_SVM_LIN}) remains valid. With the same identification steps as in Sec.~\ref{sec:ESNequiv}, we see that Eq.~(\ref{eq:EOM_SVM_BandPass}) is equivalent to the mathematical description of LI-BPNs as in Eq.~(\ref{eq:BP_LIN}). Therefore the circuit depicted in Fig.~\ref{fig:Fig1}(a) can be designed to be either an ESN or BPN, depending on whether the channel lengths vary.

\begin{figure*}[ht!]
    \centering
        \includegraphics[width=1\textwidth]{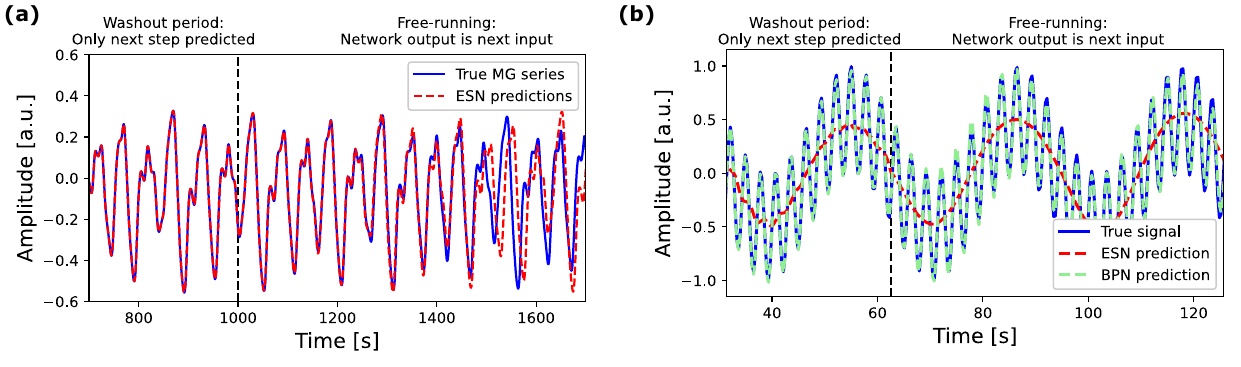}
    \caption{\label{fig:Fig2}\textbf{(a)} Mackey-Glass series predictions from a physical LI-ESN circuit containing iontronic conical channel memristors\cite{Kamsma2023IontronicMemristors}. For $t<1000$, only the next step is predicted (with time stepsize of $\delta=1~s$), for $t>1000$ the network receives no further input from the true time series and uses its own predicted output as input for the next step. Predicting 84 steps ahead yielded a RMSE of $\text{RMSE}_{84}\approx0.001$, averaged over 20 network initializations, comparable to earlier results using LI-ESNs \cite{Jaeger2010The1} and outperforming several neural network based approaches \cite{Lopez-Caraballo2016Mackey-GlassNetwork,Chen2006Time-seriesNetwork}. \textbf{(b)} Harmonic time series predictions of an LI-ESN (red) and LI-BPN (green) containing 12 iontronic SVMs, showing that LI-ESNs struggle with the variations in signal timescales.}
\end{figure*}

\section{Time series analysis tasks}
\subsection{Mackey-Glass time series}
To reproduce some of the known capabilities of LI-ESNs, and to compare to the time series prediction performance of other methods, we use our iontronic SVM based circuit to predict the synthetic Mackey-Glass \cite{Mackey1977OscillationSystems} time series $P(t)$. This time series is one of the most common generated datasets to test ESNs on \cite{Soltani2023EchoReview}, generated by
\begin{align}\label{eq:MG}
    \dfrac{\text{d}P(t)}{\text{d}t}=\frac{\beta P(t-t_{\text{delay}})}{\theta+P(t-t_{\text{delay}})^{n}}-\gamma P(t),
\end{align}
where we use $\beta=0.2$, $\theta=1$, $\gamma=0.1$, $n=10$, and $t_{\text{delay}}=17$. The first 17 time steps are randomly generated values in the range $[-1,1)$. For the aforementioned parameters, Eq.~(\ref{eq:MG}) is known to feature a chaotic attractor \cite{DoyneFarmer1982ChaoticSystem}. As in Ref. \cite{Jaeger2010The1}, Eq.~(\ref{eq:MG}) is then rescaled $P(t)\mapsto \tanh(P(t)-1)$ such that $P(t)\in[-1,1]$ for all $t$.

A reservoir was used with parameters inspired by Ref.~\cite{Jaeger2010The1} of $K=1$, $N=400$, and $L=1$, a network sparsity of 0.75, spectral radius of 0.95, $c=2.27~\text{s}$, $\delta=1~\text{s}$, a leaking rate of $a=0.95$, and an input scaling of $s^{\text{in}}=0.45~\text{V}$. The output matrix $\mathbf{W}^{\text{out}}$ was trained on a test set of length 3,000 s using ridge regression \cite{Hoerl1970RidgeProblems} discarding the first 100 s as a washout. Testing was done on a newly generated Mackey-Glass series with the same parameters, but a different random initialisation of the first 17 steps. We distinguish between a washout period, where only the next time (1 s) step needs is predicted with the true signal as input, and free-running classification, where the network receives its own output as input for the next time step while receiving no information from the true signal. The resulting washout period ($t<t_{\text{free}}=1000~\text{s}$) and free-running ($t>t_{\text{free}}=1000~\text{s}$) predictions are shown in Fig.~\ref{fig:Fig2}(a).

In Fig.~\ref{fig:Fig2}(a) we see that the network with $N=400$ is able to accurately predict the Mackey-Glass series for several 100 s. To quantify this performance and to compare it to previous results, we calculate the normalised root mean squared error (NRMSE) of the prediction $\hat{P}_i(t_{\text{free}}+84)$ with the true value $P_i(t_{\text{free}}+84)$. This entails that we compare the output of the network after it received its own prediction as input for 84 steps (i.e.\ 84 s into free-running mode), which we average over $T=20$ different random initializations of the LI-ESN, according to 

\begin{align*}
    \text{NRMSE}_{84}=\sqrt{\sum_{i=1}^{T=20}\frac{\big(\hat{P}_i(t_{\text{free}}+84)-P_i(t_{\text{free}}+84)\big)^2}{\sigma^2T}}, 
\end{align*}
with $\sigma^2\approx 0.05$ the variance of the input data.

With our fully physically realisable circuit we find $\text{NRMSE}_{84}\approx0.008$ on the test series. This outperforms other approaches, such as self-organizing feature map models reaching $\text{NRMSE}_{84}\approx 0.022$ for a comparable training set size \cite{Vesanto1997UsingPrediction} (which is the best performance in survey Ref.~ \cite{Gers2001ApplyingApproaches}), and various neural network based approaches \cite{Lopez-Caraballo2016Mackey-GlassNetwork,Chen2006Time-seriesNetwork,Ustundag2020High-PerformanceNetworks} that achieve $\text{NRMSE}_{84}\gtrsim 0.1$ (converting from RMSE to NRMSE assuming similar $\sigma^2$). This performance in and of itself is not surprising, as LI-ESNs with 400 nodes have long been shown to be capable of this \cite{Jaeger2010The1}. Therefore, recreating this using the physical conical channel SVM equations supports our claim of this work that our \emph{physical} circuit is equivalent to these ESNs and thereby to all their capabilities.

Notably, in both predictions shown in Fig.~\ref{fig:Fig2} we use the physical steady-state conductance $g_{\infty}(V)$ as activation function \cite{Kamsma2023IontronicMemristors}. Moreover, we are able to translate the parameters $a$ and $c$ to the physical length of the memristors, which for the results of Fig.~\ref{fig:Fig2}(a), with diffusion coefficient $D=1$ $\mu$m$^2$ms$^{-1}$, would be $L=169\text{ }\mu$m. Otherwise, the remaining parameters (e.g.\ channel radius, salt concentration, surface charge, etc.) are as briefly reported in Sec.~\ref{sec:SVM} and in full detail in Appendix \ref{sec:appendix_cone}.

\begin{figure*}[ht!]
    \centering
        \includegraphics[width=1\textwidth]{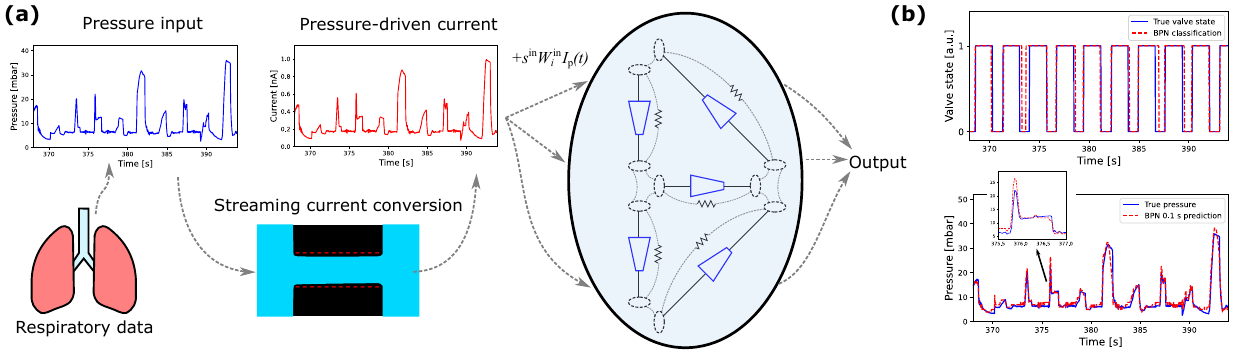}
    \caption{\label{fig:Fig3}\textbf{(a)} Schematic depiction of how the applied (airway) pressure (blue graph) drives an electric streaming current (red graph) through a cylindrical microfluidic channel (bottom) \cite{VanDerHeyden2005StreamingChannel,Werkhoven2020CoupledSystems}. This ionic current is then converted to voltage updates in the terminals of the iontronic SVMs, similar to how the ionic currents through the memristors are coupled to the neighbouring nodes. The connections in this schematic are for illustrative purposes only and do not represent a specific circuit topology used in this work. \textbf{(b)} Results of a classification task (top graph) of whether the expiratory valve (i.e.\ the valve that lets air out) is open or closed, using a small LI-BPN consisting of 8 iontronic SVMs. Additionally, a prediction task (bottom graph) of what the pressure will be 0.1 s in the future using a LI-BPN containing 200 iontronic SVMs. The inset shows that <0.1 s variations in the pressure are predicted by the network.}
\end{figure*}

\subsection{Band-pass network for multi-frequency signals}
All nodes in LI-ESNs feature the same universal relaxation timescale $c$. Therefore, LI-ESNs can struggle with inputs that incorporate signal timescales of different magnitudes \cite{Wyffels2008Band-passComputing}, whereas LI-BPNs have an inductive bias toward multi-frequency signals due to their heterogeneous timescales \cite{Wyffels2008Band-passComputing,Holzmann2010EchoReadout}. This increase in model capacity can straightforwardly be physically embedded by varying the length of the iontronic SVMs, as discussed in Sec.~\ref{sec:SVM}. To demonstrate this, we compare small $N=12$ LI-ESNs and LI-BPNs. The hyperparameters of both are optimized with the \textit{optuna framework} (version 4.4.0) \cite{Akiba2019Optuna:Framework} using the default Tree-structured Parzen Estimator algorithm, optimization code with all details is available online \cite{pyontronics}. The optimizations were for predicting a simple harmonic function that features oscillations on a $\sim1~s$ and $\sim10~s$ scale given by
\begin{align*}
    y_{\text{har}}(t)=\sin(t)\cos(1.2t),
\end{align*}
which is discretised with time step sizes of $\delta=\pi/10$ s. The networks were trained on a domain of $[0,80\pi]$. Firstly, we consider a small LI-ESN of $K=1$, $N=12$, and $L=1$, a network sparsity of 0.67, a timescale of $c=1.87$ s, a leaking rate of $a=0.44$, a spectral radius of 0.32, and an input scaling of $s^{\text{in}}=0.26~\text{V}$. The output matrix is again trained using ridge regression \cite{Hoerl1970RidgeProblems} with a regularization of $4\cdot10^{-5}$. All results presented below did not appear to be sensitive to the precise hyperparameters.

The resulting ESN predictions can be seen in the red curve of Fig.~\ref{fig:Fig2}(b), where it is clear that the LI-ESN is not able to capture the shorter timescale oscillations. Shortening the timescale of the ESN did allow it to also predict also these shorter oscillations during the washout period (i.e.\ with the true signal as input), but then the network quickly diverged in free-running mode, leading to a worse overall performance. Since we only model 12 SVM nodes with weights that are randomly generated, there is significant variability between different initializations. However, these observations were robust and the LI-ESN consistently performed poorly on predicting the time series, as we will quantify below.

Converting our LI-ESN to a similarly sized LI-BPN significantly improves the performance. In this case, timescales are drawn from a normal distribution $\mathcal{C}_i\sim N(\mu_{c},\sigma_{c}^2)$, with $\mu_c=2.79$ s and $\sigma_c=9.9$ s, such that the timescale of each SVM node is set by $c_i=\text{max}(\mu_{c}/5,\mathcal{C}_i)$ to ensure all timescales are positive. More sophisticated methods of choosing the timescales can be considered, but for now this simple approach is sufficient to demonstrate the benefits that LI-BPNs provide. Furthermore, the BPN features a network sparsity of 0.35, a leaking rate of $a=0.86$, a spectral radius of 0.76, an input scaling of $s^{\text{in}}=0.21~\text{V}$, and a regularization for ridge regression of $1\cdot10^{-6}$.

As shown in the green curve of Fig.~\ref{fig:Fig2}(b), the resulting LI-BPN can successfully predict the time series with only 12 nodes. Again, there is some variability between different initiations of the small network, but the improved performance was consistent. The RMSE of predicting the free-running domain shown in Fig.~\ref{fig:Fig2}(b) (for $[20,\pi,40\pi]$), averaged over 100 different ESN and BPN initiations, shows that the LI-BPN features a factor 3 lower RMSE ($\approx 0.04$) than the LI-ESN ($\approx0.13$). Quantitatively this seems like a somewhat marginal difference, but this is indicative of the qualitative observation that the higher frequency oscillations are not captured by the ESN circuit. We stress that physically, this conversion from an ESN to BPN is straightforward, as the variability in timescales can be realized on-chip through varying the channel lengths.

\subsection{Airway pressure as direct physical input}\label{sec:pressure}
Thus far we have provided demonstrations of analysing some synthetic time series with LI-ESNs and LI-BPNs containing iontronic SVMs. Here we will consider measurements of ventilator pressures that were designed to accurately mimic the airway pressures present in lungs during breathing \cite{ventilator-pressure-prediction}. For this task we will leverage two useful properties of iontronic SVMs, \textit{(i)} their tunable timescales that coincide with timescales of natural or biological origin, and \textit{(ii)} the intrinsic responsiveness of iontronic systems to pressure inputs.

As schematically shown in the blue curve of Fig.~\ref{fig:Fig3}(a), the input pressure is applied at one end of a cylindrical microfluidic channel Fig.~\ref{fig:Fig3}(a, bottom) of length $L=200$ $\mu$m and radius $R=25$ $\mu$m, carrying a typical surface potential of $\psi_0=-40~\text{mV}$. These channels feature a coupling between pressure and electric (ionic) current, specifically the resulting pressure drop $\Delta p(t)$ drives a so-called electric \textit{streaming current} (red curve of Fig.~\ref{fig:Fig3}(a)) given by\cite{Werkhoven2020CoupledSystems}
\begin{align*}
    I_{\text{p}}(t)=\pi R^2\frac{\epsilon\psi_0}{\eta}\frac{\Delta p(t)}{L},
\end{align*}
with $\eta$ and $\epsilon$ the shear viscosity and electric permittivity of water. The length $L$ and radius $R$ of the channel are chosen such that for the typical biological pressure signal amplitude of $\sim10$ mbar, the streaming current $I_{\text{p}}(t)$ will be of order $\sim1$ nA. This is a current strength that can be reliably measured for single channels driven by pressure \cite{VanDerHeyden2005StreamingChannel,Jubin2018DramaticNanopores}. The ionic streaming current becomes the input $u(t)=s^{\text{in}}I_{\text{p}}(t)$. Similar to how the ionic currents through the SVMs are converted to voltage contribution updates within the terminals, the streaming current can now be directly coupled to the SVMs such that the voltage contribution $s^{\text{in}}I_{\text{p}}(t)W_{i}^{\text{in}}$ of the inputs is as per Eq.~(\ref{eq:VoltRule}).

Using the ventilator pressure data\cite{ventilator-pressure-prediction}, an example of which is shown (blue) in Fig.~\ref{fig:Fig3}(a),  we perform two tasks. First, the easier task of classifying whether the expiratory valve (i.e.\ the valve that lets air out) is open or closed, and then the harder task of predicting 3 steps ($\approx 0.1$ s) ahead, i.e.\ pressure is applied as input and the goal is to predict step $n+3$ when at step $n$. The data was split into two parts for training and testing, an 80,000 step  (i.e.\ ~2700 s) segment for training, and a different 20,000 step (i.e.\ ~680 s) segment for testing (in both instances discarding the first 1000 steps as washout). Hyperparameters were optimized for analysing ventilator pressure data again with optuna framework (version 4.4.0) \cite{Akiba2019Optuna:Framework}, using the default Tree-structured Parzen Estimator algorithm, optimization code with all details is available online \cite{pyontronics}. The resulting LI-BPNs feature a network sparsity of 0.017, time step $\delta=0.034035$ s (matching with the data), timescale parameters $\mu_{c}=0.27~\text{s}$ s and $\sigma_{c}=1.89~\text{s}$, input scaling $s^{\text{in}}=0.11~\text{V/nA}$, a leaking rate of $a=0.98$, and a regularization of $1.69\cdot10^{-6}$.

In the top graph of Fig.~\ref{fig:Fig3}(b) we show the valve classifications in red, compared to the true value in blue, corresponding to the input depicted in Fig.~\ref{fig:Fig3}(a). For these classifications a small LI-BPN circuit was modelled with only $N=7$ iontronic SVMs, achieving $\sim91\%$ on the test data accuracy averaged over 20 initialisations. For comparison, we carried out the same task with a linear autoregression model of order 7, such that the number of fit-parameters matches the LI-BPN, achieving a lower accuracy of $\sim 82\%$. This is partially because such a model only uses a slim window of the last 7 time steps for its classification. Spreading out this window through subsampling by using the values $8m$ steps back, with $m\in\{1,2,...,7\}$, improves accuracy to $90\%$, almost matching the LI-BPN. However, it is not directly clear how a simple hardware implementation of such a (pressure-driven) autoregression model would be achieved, especially in the more complicated subsampling approach.

In the bottom graph of Fig.~\ref{fig:Fig3}(b) we show the pressure predictions using a larger LI-BPN circuit containing $N=200$ SVMs (otherwise the same parameters). Each point in the red graph is a prediction of 0.1 s ahead, achieving an RMSE of $\approx3.0$ mbar (measured over the full test data length).  This is a slight improvement over the RMSE of $\approx3.2$ mbar achieved with a linear autoregression model of order 200, where we again note that it is not directly clear how such a (pressure-driven) autoregression model can be straightforwardly directly physically implemented like the LI-BPN. Although most individual pressure waves are longer than the prediction window of 0.1 s, there are certainly $<0.1~s$ features within each pressure wave that are still correctly predicted by the network, as shown in the inset in the bottom graph of Fig.~\ref{fig:Fig3}(b). We note that this task is especially difficult because normally multiple parameters accompany each individual waveform \cite{ventilator-pressure-prediction}, whereas here we solely provide the pressure as input.

\section{Discussion and conclusion}
In summary, we proposed a physical circuit design that exhibits a one-to-one correspondence to the well-established mathematical description of the reservoir computing frameworks of Leaky Integrator Echo State and Band-pass Networks\cite{Jaeger2002AdaptiveNetworks,Jaeger2007OptimizationNeurons,Jaeger2010The1,Yildiz2012Re-visitingProperty,Lukosevicius2012ANetworks,Whiteaker2021LeakySets,Lin2024AComputing,Sun2024AApplication,Wyffels2008Band-passComputing}. This circuit incorporates fluidic iontronic memristors \cite{Noy2023FluidDevices,Noy2023NanofluidicSplash,Hou2023LearningNanofluidics,Yu2023BioinspiredComputing,Xie2022PerspectiveApplication}, whose voltage dependent conductance and conductance memory facilitate the activation function and dynamics of the nodes, respectively. The terminals at either end of the memristors feature fixed peripheral circuits that convert locally measured currents to voltages, forming a physical realization of the matrix coupling between nodes. By solving the physical equations of microfluidic conical channel memristors \cite{Kamsma2023IontronicMemristors} within such a circuit, we successfully analyze several signals such as the Mackey-Glass time series, supporting our claim that such a circuit is a physical manifestation of LI-ESNs and LI-BPNs. 

A desirably property of these iontronic memristors is the dependence of their conductance memory timescale on the channel length, i,e,\ each memristor can individually be designed to feature a certain relaxation time. This corresponds to going from a LI-ESN to a LI-BPN, which are known to perform significantly better on inputs that feature multiple timescales \cite{Wyffels2008Band-passComputing}, demonstrated here by showing a circuit with channels of differing lengths, i.e.\ a physical LI-BPN circuit, performs significantly better on predicting a harmonic time series with features across different timescales than a physical LI-ESN. 

Lastly, we leverage another unique property of microfluidic devices, their intrinsic responsiveness to applied pressures. Applied pressures can drive electric currents through microfluidic channels, allowing a pressure signal to be converted to an ionic current. By coupling this to the existing current-to-voltage conversion in the terminals through an additional input channel, the pressure is converted to the network's input without any intervention or interaction required from outside the network. We demonstrated this by classifying and predicting features of simulated biophysically realistic data of airway pressures during breathing \cite{ventilator-pressure-prediction}.

Some functionality of our proposed circuit design lies in the peripheral circuitry that connect the terminals of the memristors, namely converting currents of neighbouring memristors to updated voltages. Current-to-voltage converting peripheral circuits are common within neuromorphics \cite{Li2018EfficientNetworks}, supporting that such circuitry can efficiently be implemented. We ignored device noise in this study, assuming that the currents through the memristors can be accurately and reliably measured. Interestingly, LI-ESNs have also been theoretically studied with noise \cite{Jaeger2002AdaptiveNetworks,Jaeger2007OptimizationNeurons,Sun2024AApplication}, so both from a theoretical and from a physical implementation perspective this is an interesting direction to study next. Additionally, analysing more real-world data, such as other biological (pressure) sources or chemical signals \cite{Xiong2023NeuromorphicMemristor,Ling2024Single-PoreCombinations,Kamsma2025ChemicallyApplications}, would be a relevant expansion. Lastly, optimising the circuit design for specific tasks could be of interest, as our focus here was primarily on establishing the correspondence between our physical circuit design and LI-ESNs or LI-BPNs, not on optimization.

With power consumptions of iontronic memristive devices as low as order 10 fW per channel \cite{Shi2023UltralowMemristor}, the overall circuit could potentially operate at very low power, where we leave a more thorough estimate for future work. The energy usage will also depend on the peripheral circuit, which can be a significant contributor to power consumption \cite{Zahedi2020EfficientCim}, and on its ability to integrate such low power (sub)-nanoscale devices. Input and output actions, performed by standard matrix multiplications, could also be physically realized using crossbar arrays implemented with (ionic) devices \cite{Liu2025Resistance-RestorableChip,VanDeBurgt2018OrganicComputing,Kazemzadeh2025AllArray,Xu2025Angstrom-Scale-ChannelComputing,Zhang2022AdaptiveConductor,Hu2023AnComputing,VanDeBurgt2017AComputing}, further supporting our design as a fully physically realisable low power Echo State or Band-pass Network circuit.

In conclusion, our proposed iontronic memristor based physical circuit design is theoretically equivalent to the well-established reservoir computing methods of Leaky Integrator Echo State and Band-pass Networks. This is supported by performing several time series prediction and classification tasks. The fluidic devices do not necessarily need to be integrated in a fully fluidic circuit, thereby circumventing existing challenges of manufacturing integrated fully fluidic chips \cite{FraimanTowardComponents}. Notably, airway pressure signals were used as inputs, leveraging iontronics' intrinsic pressure responsiveness and natural timescales matching biology, with pressure classification tasks achieved using as few as 7 memristors. Therefore, this work provides a design that can advance the field of iontronic (neuromorphic) computing, while exploiting some of iontronics' intrinsic properties.

\begin{acknowledgments}
We thank Dr.\ Jaehyun Kim and Prof.\ Jungyul Park for providing us with the Colloid Based channel schematic shown in Fig.~\ref{fig:Fig1}(c).
\end{acknowledgments}

\appendix

\section{Conical channel conductance}\label{sec:appendix_cone}
This Appendix is a summary of the results described in Ref.\cite{Kamsma2023IontronicMemristors}. We consider an azimuthally symmetric single conical channel, schematically depicted in Fig.~\ref{fig:Fig1}(b, top left), of length $L$ (given for each specific network in the main text) with the central axis at radial coordinate $r=0$ and a radius described by $R(x)=R_{\mathrm{b}}-x\Delta R/L$ for $x\in\left[0,L\right]$ where $R_{\mathrm{b}}=200\text{ nm}$ is the base radius at $x=0$ and $R_{\mathrm{t}}=R_{\mathrm{b}}-\Delta R=50\text{ nm}$ the tip radius at $x=L\gg R_{\mathrm{b}}$. The channel connects two bulk reservoirs of an incompressible aqueous 1:1 electrolyte with viscosity $\eta=1.01\text{ mPa}\cdot\text{s}$, mass density $\rho_{\mathrm{m}}=10^3\text{ kg}\cdot\text{m}^{-3}$ and electric permittivity $\epsilon=0.71\text{ nF}\cdot\text{m}^{-1}$, at the far side of both reservoirs we impose a fixed pressure $P=P_0$ and fixed ion concentrations $\rho_{\pm}=\rho_{\mathrm{b}}=0.1\text{ mM}$. The channel wall carries a uniform surface charge density $e\sigma=-0.0015\;e\text{nm}^{-2}$, screened by an electric double layer with Debye length $\lambda_{\mathrm{D}}\approx10\text{ nm}$, resulting in an electric surface potential of $\psi_0\approx -10\text{ mV}$. The ions have concentrations $\rho_\pm(x,r)$, diffusion coefficients $D_{\pm}=D=1\text{ }\mu\text{m}^2\text{ms}^{-1}$, and charge $\pm e$ with $e$ the proton charge. Over the channel we impose an electric potential $V(t)$, defined as the voltage $V_{\mathrm{t}}(t)$ in the tip reservoir minus the voltage $V_{\mathrm{b}}(t)$in the base reservoir.

The steady-state conductance of a conical channel depends on the voltage-dependent radially averaged salt concentration profile $\overline{\rho}_{\mathrm{s}}(x,V)=2\int_{0}^{R(x)}r(\rho_{\mathrm{+}}(x,r)+\rho_{\mathrm{-}}(x,r))\mathrm{d} r/R(x)^2$ that exhibits salt concentration polarisation upon an applied voltage. The consequent voltage-dependent steady-state channel conductance is described by \cite{Boon2022Pressure-sensitiveGeometry}
\begin{equation}\label{eq:rhos}
    \begin{aligned}
    \frac{g_{\infty}(V)}{g_0}=&\int_0^L \overline{\rho}_{\mathrm{s}}(x,V) \mathrm{d} x/(2\rho_{\rm{b}}L)\\
    =&1+\Delta g\int_0^{L}\left[\frac{x}{L}\frac{R_{\mathrm{t}}}{R(x)}-\frac{e^{\text{Pe}(V)\frac{x}{L}\frac{ R_{\mathrm{t}}^2}{R_{\mathrm{b}}R(x)}}-1}{e^{\text{Pe}(V)\frac{R_{\mathrm{t}}}{R_{\mathrm{b}}}}-1}\right]\mathrm{d} x/L,
    \end{aligned}
\end{equation}
where $g_0=(\pi R_{\mathrm{t}} R_{\mathrm{b}}/L)(2\rho_{\rm{b}}e^2D/k_{\mathrm{B}}T)$, $\text{Pe}(V)=Q(V)L/(D\pi R_{\mathrm{t}}^2)$ the P\'{e}clet number at the narrow end, $Q(V)=-\pi R_{\mathrm{t}}R_{\mathrm{b}}\epsilon\psi_0/(\eta L)V$ the volumetric fluid flow  through the channel, and $\Delta g\equiv-e\Delta R\eta\sigma D/(\rho_{\mathrm{b}}R_{\mathrm{b}}R_{\mathrm{t}}\epsilon\psi_0k_{\mathrm{B}}T)$.

The dynamic (dimensional) conductance $g(t)$ is well described by \cite{Kamsma2023IontronicMemristors}
\begin{align}
    \dfrac{\mathrm{d} g(t)}{\mathrm{d} t}=\frac{g_{\infty}(V(t))-g(t)}{\tau},
\end{align}
with $\tau=L^2/12D$ the typical conductance memory time of the channel.


\begin{thebibliography}{69}%
\makeatletter
\providecommand \@ifxundefined [1]{%
 \@ifx{#1\undefined}
}%
\providecommand \@ifnum [1]{%
 \ifnum #1\expandafter \@firstoftwo
 \else \expandafter \@secondoftwo
 \fi
}%
\providecommand \@ifx [1]{%
 \ifx #1\expandafter \@firstoftwo
 \else \expandafter \@secondoftwo
 \fi
}%
\providecommand \natexlab [1]{#1}%
\providecommand \enquote  [1]{``#1''}%
\providecommand \bibnamefont  [1]{#1}%
\providecommand \bibfnamefont [1]{#1}%
\providecommand \citenamefont [1]{#1}%
\providecommand \href@noop [0]{\@secondoftwo}%
\providecommand \href [0]{\begingroup \@sanitize@url \@href}%
\providecommand \@href[1]{\@@startlink{#1}\@@href}%
\providecommand \@@href[1]{\endgroup#1\@@endlink}%
\providecommand \@sanitize@url [0]{\catcode `\\12\catcode `\$12\catcode `\&12\catcode `\#12\catcode `\^12\catcode `\_12\catcode `\%12\relax}%
\providecommand \@@startlink[1]{}%
\providecommand \@@endlink[0]{}%
\providecommand \url  [0]{\begingroup\@sanitize@url \@url }%
\providecommand \@url [1]{\endgroup\@href {#1}{\urlprefix }}%
\providecommand \urlprefix  [0]{URL }%
\providecommand \Eprint [0]{\href }%
\providecommand \doibase [0]{https://doi.org/}%
\providecommand \selectlanguage [0]{\@gobble}%
\providecommand \bibinfo  [0]{\@secondoftwo}%
\providecommand \bibfield  [0]{\@secondoftwo}%
\providecommand \translation [1]{[#1]}%
\providecommand \BibitemOpen [0]{}%
\providecommand \bibitemStop [0]{}%
\providecommand \bibitemNoStop [0]{.\EOS\space}%
\providecommand \EOS [0]{\spacefactor3000\relax}%
\providecommand \BibitemShut  [1]{\csname bibitem#1\endcsname}%
\let\auto@bib@innerbib\@empty
\bibitem [{\citenamefont {Nakajima}\ and\ \citenamefont {Fischer}(2021)}]{Nakajima2021NaturalApplications}%
  \BibitemOpen
  \bibfield  {author} {\bibinfo {author} {\bibfnamefont {K.}~\bibnamefont {Nakajima}}\ and\ \bibinfo {author} {\bibfnamefont {I.}~\bibnamefont {Fischer}},\ }\href {http://www.springer.com/series/4190} {\emph {\bibinfo {title} {{Natural Computing Series Reservoir Computing Theory, Physical Implementations, and Applications}}}},\ \bibinfo {type} {Tech. Rep.}\ (\bibinfo {year} {2021})\BibitemShut {NoStop}%
\bibitem [{\citenamefont {Cucchi}\ \emph {et~al.}(2022)\citenamefont {Cucchi}, \citenamefont {Abreu}, \citenamefont {Ciccone}, \citenamefont {Brunner},\ and\ \citenamefont {Kleemann}}]{Cucchi2022Hands-onImplementation}%
  \BibitemOpen
  \bibfield  {author} {\bibinfo {author} {\bibfnamefont {M.}~\bibnamefont {Cucchi}}, \bibinfo {author} {\bibfnamefont {S.}~\bibnamefont {Abreu}}, \bibinfo {author} {\bibfnamefont {G.}~\bibnamefont {Ciccone}}, \bibinfo {author} {\bibfnamefont {D.}~\bibnamefont {Brunner}},\ and\ \bibinfo {author} {\bibfnamefont {H.}~\bibnamefont {Kleemann}},\ }\bibfield  {title} {\bibinfo {title} {{Hands-on reservoir computing: a tutorial for practical implementation}},\ }\bibfield  {journal} {\bibinfo  {journal} {Neuromorphic Computing and Engineering}\ }\textbf {\bibinfo {volume} {2}},\ \href {https://doi.org/10.1088/2634-4386/ac7db7} {10.1088/2634-4386/ac7db7} (\bibinfo {year} {2022})\BibitemShut {NoStop}%
\bibitem [{\citenamefont {Liang}\ \emph {et~al.}(2024)\citenamefont {Liang}, \citenamefont {Tang}, \citenamefont {Zhong}, \citenamefont {Gao}, \citenamefont {Qian},\ and\ \citenamefont {Wu}}]{Liang2024PhysicalElectronics}%
  \BibitemOpen
  \bibfield  {author} {\bibinfo {author} {\bibfnamefont {X.}~\bibnamefont {Liang}}, \bibinfo {author} {\bibfnamefont {J.}~\bibnamefont {Tang}}, \bibinfo {author} {\bibfnamefont {Y.}~\bibnamefont {Zhong}}, \bibinfo {author} {\bibfnamefont {B.}~\bibnamefont {Gao}}, \bibinfo {author} {\bibfnamefont {H.}~\bibnamefont {Qian}},\ and\ \bibinfo {author} {\bibfnamefont {H.}~\bibnamefont {Wu}},\ }\bibfield  {title} {\bibinfo {title} {{Physical reservoir computing with emerging electronics}},\ }\href {https://doi.org/10.1038/s41928-024-01133-z} {\bibfield  {journal} {\bibinfo  {journal} {Nature Electronics 2024 7:3}\ }\textbf {\bibinfo {volume} {7}},\ \bibinfo {pages} {193} (\bibinfo {year} {2024})}\BibitemShut {NoStop}%
\bibitem [{\citenamefont {Noy}\ \emph {et~al.}(2023)\citenamefont {Noy}, \citenamefont {Li},\ and\ \citenamefont {Darling}}]{Noy2023FluidDevices}%
  \BibitemOpen
  \bibfield  {author} {\bibinfo {author} {\bibfnamefont {A.}~\bibnamefont {Noy}}, \bibinfo {author} {\bibfnamefont {Z.}~\bibnamefont {Li}},\ and\ \bibinfo {author} {\bibfnamefont {S.~B.}\ \bibnamefont {Darling}},\ }\bibfield  {title} {\bibinfo {title} {{Fluid learning: Mimicking brain computing with neuromorphic nanofluidic devices}},\ }\href {https://doi.org/10.1016/J.NANTOD.2023.102043} {\bibfield  {journal} {\bibinfo  {journal} {Nano Today}\ }\textbf {\bibinfo {volume} {53}},\ \bibinfo {pages} {102043} (\bibinfo {year} {2023})}\BibitemShut {NoStop}%
\bibitem [{\citenamefont {Noy}\ and\ \citenamefont {Darling}(2023)}]{Noy2023NanofluidicSplash}%
  \BibitemOpen
  \bibfield  {author} {\bibinfo {author} {\bibfnamefont {A.}~\bibnamefont {Noy}}\ and\ \bibinfo {author} {\bibfnamefont {S.~B.}\ \bibnamefont {Darling}},\ }\bibfield  {title} {\bibinfo {title} {{Nanofluidic computing makes a splash}},\ }\href {https://doi.org/10.1126/science.adf6400} {\bibfield  {journal} {\bibinfo  {journal} {Science}\ }\textbf {\bibinfo {volume} {379}},\ \bibinfo {pages} {143} (\bibinfo {year} {2023})}\BibitemShut {NoStop}%
\bibitem [{\citenamefont {Hou}\ \emph {et~al.}(2023)\citenamefont {Hou}, \citenamefont {Ling}, \citenamefont {Wang}, \citenamefont {Wang}, \citenamefont {Chen}, \citenamefont {Li},\ and\ \citenamefont {Hou}}]{Hou2023LearningNanofluidics}%
  \BibitemOpen
  \bibfield  {author} {\bibinfo {author} {\bibfnamefont {Y.}~\bibnamefont {Hou}}, \bibinfo {author} {\bibfnamefont {Y.}~\bibnamefont {Ling}}, \bibinfo {author} {\bibfnamefont {Y.}~\bibnamefont {Wang}}, \bibinfo {author} {\bibfnamefont {M.}~\bibnamefont {Wang}}, \bibinfo {author} {\bibfnamefont {Y.}~\bibnamefont {Chen}}, \bibinfo {author} {\bibfnamefont {X.}~\bibnamefont {Li}},\ and\ \bibinfo {author} {\bibfnamefont {X.}~\bibnamefont {Hou}},\ }\bibfield  {title} {\bibinfo {title} {{Learning from the Brain: Bioinspired Nanofluidics}},\ }\href {https://doi.org/10.1021/acs.jpclett.2c03930} {\bibfield  {journal} {\bibinfo  {journal} {Journal of Physical Chemistry Letters}\ }\textbf {\bibinfo {volume} {14}},\ \bibinfo {pages} {2891} (\bibinfo {year} {2023})}\BibitemShut {NoStop}%
\bibitem [{\citenamefont {Yu}\ \emph {et~al.}(2023)\citenamefont {Yu}, \citenamefont {Li}, \citenamefont {Luo}, \citenamefont {Lei}, \citenamefont {Wang}, \citenamefont {Hou}, \citenamefont {Wang},\ and\ \citenamefont {Hou}}]{Yu2023BioinspiredComputing}%
  \BibitemOpen
  \bibfield  {author} {\bibinfo {author} {\bibfnamefont {L.}~\bibnamefont {Yu}}, \bibinfo {author} {\bibfnamefont {X.}~\bibnamefont {Li}}, \bibinfo {author} {\bibfnamefont {C.}~\bibnamefont {Luo}}, \bibinfo {author} {\bibfnamefont {Z.}~\bibnamefont {Lei}}, \bibinfo {author} {\bibfnamefont {Y.}~\bibnamefont {Wang}}, \bibinfo {author} {\bibfnamefont {Y.}~\bibnamefont {Hou}}, \bibinfo {author} {\bibfnamefont {M.}~\bibnamefont {Wang}},\ and\ \bibinfo {author} {\bibfnamefont {X.}~\bibnamefont {Hou}},\ }\bibfield  {title} {\bibinfo {title} {{Bioinspired nanofluidic iontronics for brain-like computing}},\ }\href {https://doi.org/10.1007/s12274-023-5900-y} {\bibfield  {journal} {\bibinfo  {journal} {Nano Research}\ ,\ \bibinfo {pages} {1}} (\bibinfo {year} {2023})}\BibitemShut {NoStop}%
\bibitem [{\citenamefont {Xie}\ \emph {et~al.}(2022)\citenamefont {Xie}, \citenamefont {Xiong}, \citenamefont {Li}, \citenamefont {Gao}, \citenamefont {Zong}, \citenamefont {Liu},\ and\ \citenamefont {Yu}}]{Xie2022PerspectiveApplication}%
  \BibitemOpen
  \bibfield  {author} {\bibinfo {author} {\bibfnamefont {B.}~\bibnamefont {Xie}}, \bibinfo {author} {\bibfnamefont {T.}~\bibnamefont {Xiong}}, \bibinfo {author} {\bibfnamefont {W.}~\bibnamefont {Li}}, \bibinfo {author} {\bibfnamefont {T.}~\bibnamefont {Gao}}, \bibinfo {author} {\bibfnamefont {J.}~\bibnamefont {Zong}}, \bibinfo {author} {\bibfnamefont {Y.}~\bibnamefont {Liu}},\ and\ \bibinfo {author} {\bibfnamefont {P.}~\bibnamefont {Yu}},\ }\bibfield  {title} {\bibinfo {title} {{Perspective on Nanofluidic Memristors: From Mechanism to Application}},\ }\href {https://doi.org/10.1002/ASIA.202200682} {\bibfield  {journal} {\bibinfo  {journal} {Chemistry - An Asian Journal}\ }\textbf {\bibinfo {volume} {17}},\ \bibinfo {pages} {e202200682} (\bibinfo {year} {2022})}\BibitemShut {NoStop}%
\bibitem [{\citenamefont {Fan}\ \emph {et~al.}(2025)\citenamefont {Fan}, \citenamefont {Shang}, \citenamefont {Yuan}, \citenamefont {Zhang},\ and\ \citenamefont {Sha}}]{Fan2025EmergingComputation}%
  \BibitemOpen
  \bibfield  {author} {\bibinfo {author} {\bibfnamefont {Q.}~\bibnamefont {Fan}}, \bibinfo {author} {\bibfnamefont {J.}~\bibnamefont {Shang}}, \bibinfo {author} {\bibfnamefont {X.}~\bibnamefont {Yuan}}, \bibinfo {author} {\bibfnamefont {Z.}~\bibnamefont {Zhang}},\ and\ \bibinfo {author} {\bibfnamefont {J.}~\bibnamefont {Sha}},\ }\bibfield  {title} {\bibinfo {title} {{Emerging Liquid-Based Memristive Devices for Neuromorphic Computation}},\ }\href {https://doi.org/10.1002/SMTD.202402218;CTYPE:STRING:JOURNAL} {\bibfield  {journal} {\bibinfo  {journal} {Small Methods}\ ,\ \bibinfo {pages} {2402218}} (\bibinfo {year} {2025})}\BibitemShut {NoStop}%
\bibitem [{\citenamefont {Suwen~Law}\ \emph {et~al.}(2025)\citenamefont {Suwen~Law}, \citenamefont {Wang}, \citenamefont {Nielsch}, \citenamefont {Abell}, \citenamefont {Bisquert}, \citenamefont {Santos},\ and\ \citenamefont {Appl~Phys}}]{SuwenLaw2025RecentComputing}%
  \BibitemOpen
  \bibfield  {author} {\bibinfo {author} {\bibfnamefont {C.}~\bibnamefont {Suwen~Law}}, \bibinfo {author} {\bibfnamefont {J.}~\bibnamefont {Wang}}, \bibinfo {author} {\bibfnamefont {K.}~\bibnamefont {Nielsch}}, \bibinfo {author} {\bibfnamefont {A.~D.}\ \bibnamefont {Abell}}, \bibinfo {author} {\bibfnamefont {J.}~\bibnamefont {Bisquert}}, \bibinfo {author} {\bibfnamefont {A.}~\bibnamefont {Santos}},\ and\ \bibinfo {author} {\bibfnamefont {J.}~\bibnamefont {Appl~Phys}},\ }\bibfield  {title} {\bibinfo {title} {{Recent advances in fluidic neuromorphic computing}},\ }\href {https://doi.org/10.1063/5.0235267} {\bibfield  {journal} {\bibinfo  {journal} {Applied Physics Reviews}\ }\textbf {\bibinfo {volume} {12}},\ \bibinfo {pages} {21309} (\bibinfo {year} {2025})}\BibitemShut {NoStop}%
\bibitem [{\citenamefont {Lv}\ \emph {et~al.}(2025)\citenamefont {Lv}, \citenamefont {Liu},\ and\ \citenamefont {Zhang}}]{Lv2025AdvancementsDesign}%
  \BibitemOpen
  \bibfield  {author} {\bibinfo {author} {\bibfnamefont {H.}~\bibnamefont {Lv}}, \bibinfo {author} {\bibfnamefont {R.}~\bibnamefont {Liu}},\ and\ \bibinfo {author} {\bibfnamefont {Y.}~\bibnamefont {Zhang}},\ }\bibfield  {title} {\bibinfo {title} {{Advancements in Fluidic Ionic Devices: Implications for Neuromorphic Integrated Circuit Design}},\ }\bibfield  {journal} {\bibinfo  {journal} {ACS Sensors}\ }\href {https://doi.org/10.1021/ACSSENSORS.5C01063} {10.1021/ACSSENSORS.5C01063} (\bibinfo {year} {2025})\BibitemShut {NoStop}%
\bibitem [{\citenamefont {Kamsma}\ \emph {et~al.}(2023{\natexlab{a}})\citenamefont {Kamsma}, \citenamefont {Kim}, \citenamefont {Kim}, \citenamefont {Boon}, \citenamefont {Spitoni}, \citenamefont {Park},\ and\ \citenamefont {van Roij}}]{Kamsma2023Brain-inspiredNanochannels}%
  \BibitemOpen
  \bibfield  {author} {\bibinfo {author} {\bibfnamefont {T.~M.}\ \bibnamefont {Kamsma}}, \bibinfo {author} {\bibfnamefont {J.}~\bibnamefont {Kim}}, \bibinfo {author} {\bibfnamefont {K.}~\bibnamefont {Kim}}, \bibinfo {author} {\bibfnamefont {W.~Q.}\ \bibnamefont {Boon}}, \bibinfo {author} {\bibfnamefont {C.}~\bibnamefont {Spitoni}}, \bibinfo {author} {\bibfnamefont {J.}~\bibnamefont {Park}},\ and\ \bibinfo {author} {\bibfnamefont {R.}~\bibnamefont {van Roij}},\ }\bibfield  {title} {\bibinfo {title} {{Brain-inspired computing with fluidic iontronic nanochannels}},\ }\href {https://doi.org/10.1073/PNAS.2320242121/SUPPL{\_}FILE/PNAS.2320242121.SAPP.PDF} {\bibfield  {journal} {\bibinfo  {journal} {Proceedings of the National Academy of Sciences}\ }\textbf {\bibinfo {volume} {121}},\ \bibinfo {pages} {e2320242121} (\bibinfo {year} {2023}{\natexlab{a}})}\BibitemShut {NoStop}%
\bibitem [{\citenamefont {Portillo}\ \emph {et~al.}(2025)\citenamefont {Portillo}, \citenamefont {Ramirez}, \citenamefont {Mafe},\ and\ \citenamefont {Cervera}}]{Portillo2025NeuromorphicDiodes}%
  \BibitemOpen
  \bibfield  {author} {\bibinfo {author} {\bibfnamefont {S.}~\bibnamefont {Portillo}}, \bibinfo {author} {\bibfnamefont {P.}~\bibnamefont {Ramirez}}, \bibinfo {author} {\bibfnamefont {S.}~\bibnamefont {Mafe}},\ and\ \bibinfo {author} {\bibfnamefont {J.}~\bibnamefont {Cervera}},\ }\bibfield  {title} {\bibinfo {title} {{Neuromorphic Reservoir Computing with Memristive Nanofluidic Diodes}},\ }\href {https://doi.org/10.1021/ACS.NANOLETT.5C00853/ASSET/IMAGES/LARGE/NL5C00853{\_}0005.JPEG} {\bibfield  {journal} {\bibinfo  {journal} {Nano Letters}\ }\textbf {\bibinfo {volume} {13}},\ \bibinfo {pages} {53} (\bibinfo {year} {2025})}\BibitemShut {NoStop}%
\bibitem [{\citenamefont {Cervera}\ \emph {et~al.}(2024)\citenamefont {Cervera}, \citenamefont {Portillo}, \citenamefont {Ramirez},\ and\ \citenamefont {Mafe}}]{Cervera2024ModelingDiodes}%
  \BibitemOpen
  \bibfield  {author} {\bibinfo {author} {\bibfnamefont {J.}~\bibnamefont {Cervera}}, \bibinfo {author} {\bibfnamefont {S.}~\bibnamefont {Portillo}}, \bibinfo {author} {\bibfnamefont {P.}~\bibnamefont {Ramirez}},\ and\ \bibinfo {author} {\bibfnamefont {S.}~\bibnamefont {Mafe}},\ }\bibfield  {title} {\bibinfo {title} {{Modeling of memory effects in nanofluidic diodes}},\ }\href {https://doi.org/10.1063/5.0204219/3283134} {\bibfield  {journal} {\bibinfo  {journal} {Physics of Fluids}\ }\textbf {\bibinfo {volume} {36}},\ \bibinfo {pages} {47129} (\bibinfo {year} {2024})}\BibitemShut {NoStop}%
\bibitem [{\citenamefont {Kamsma}\ \emph {et~al.}(2023{\natexlab{b}})\citenamefont {Kamsma}, \citenamefont {Boon}, \citenamefont {Ter~Rele}, \citenamefont {Spitoni},\ and\ \citenamefont {Van~Roij}}]{Kamsma2023IontronicMemristors}%
  \BibitemOpen
  \bibfield  {author} {\bibinfo {author} {\bibfnamefont {T.~M.}\ \bibnamefont {Kamsma}}, \bibinfo {author} {\bibfnamefont {W.~Q.}\ \bibnamefont {Boon}}, \bibinfo {author} {\bibfnamefont {T.}~\bibnamefont {Ter~Rele}}, \bibinfo {author} {\bibfnamefont {C.}~\bibnamefont {Spitoni}},\ and\ \bibinfo {author} {\bibfnamefont {R.}~\bibnamefont {Van~Roij}},\ }\bibfield  {title} {\bibinfo {title} {{Iontronic Neuromorphic Signaling with Conical Microfluidic Memristors}},\ }\href {https://doi.org/10.1103/PhysRevLett.130.268401} {\bibfield  {journal} {\bibinfo  {journal} {Physical Review Letters}\ }\textbf {\bibinfo {volume} {130}},\ \bibinfo {pages} {268401} (\bibinfo {year} {2023}{\natexlab{b}})}\BibitemShut {NoStop}%
\bibitem [{\citenamefont {Zhang}\ \emph {et~al.}(2024)\citenamefont {Zhang}, \citenamefont {Sabbagh}, \citenamefont {Chen},\ and\ \citenamefont {Yossifon}}]{Zhang2024GeometricallySystems}%
  \BibitemOpen
  \bibfield  {author} {\bibinfo {author} {\bibfnamefont {Z.}~\bibnamefont {Zhang}}, \bibinfo {author} {\bibfnamefont {B.}~\bibnamefont {Sabbagh}}, \bibinfo {author} {\bibfnamefont {Y.}~\bibnamefont {Chen}},\ and\ \bibinfo {author} {\bibfnamefont {G.}~\bibnamefont {Yossifon}},\ }\bibfield  {title} {\bibinfo {title} {{Geometrically Scalable Iontronic Memristors: Employing Bipolar Polyelectrolyte Gels for Neuromorphic Systems}},\ }\bibfield  {journal} {\bibinfo  {journal} {ACS Nano}\ }\href {https://doi.org/10.1021/ACSNANO.4C01730} {10.1021/ACSNANO.4C01730} (\bibinfo {year} {2024})\BibitemShut {NoStop}%
\bibitem [{\citenamefont {Kamsma}\ \emph {et~al.}(2024{\natexlab{a}})\citenamefont {Kamsma}, \citenamefont {Rossing}, \citenamefont {Spitoni},\ and\ \citenamefont {Roij}}]{Kamsma2024AdvancedCircuit}%
  \BibitemOpen
  \bibfield  {author} {\bibinfo {author} {\bibfnamefont {T.~M.}\ \bibnamefont {Kamsma}}, \bibinfo {author} {\bibfnamefont {E.~A.}\ \bibnamefont {Rossing}}, \bibinfo {author} {\bibfnamefont {C.}~\bibnamefont {Spitoni}},\ and\ \bibinfo {author} {\bibfnamefont {R.~v.}\ \bibnamefont {Roij}},\ }\bibfield  {title} {\bibinfo {title} {{Advanced iontronic spiking modes with multiscale diffusive dynamics in a fluidic circuit}},\ }\href {https://doi.org/10.1088/2634-4386/AD40CA} {\bibfield  {journal} {\bibinfo  {journal} {Neuromorphic Computing and Engineering}\ }\textbf {\bibinfo {volume} {4}},\ \bibinfo {pages} {024003} (\bibinfo {year} {2024}{\natexlab{a}})}\BibitemShut {NoStop}%
\bibitem [{\citenamefont {Chicca}\ and\ \citenamefont {Indiveri}(2020)}]{Chicca2020ASystems}%
  \BibitemOpen
  \bibfield  {author} {\bibinfo {author} {\bibfnamefont {E.}~\bibnamefont {Chicca}}\ and\ \bibinfo {author} {\bibfnamefont {G.}~\bibnamefont {Indiveri}},\ }\bibfield  {title} {\bibinfo {title} {{A recipe for creating ideal hybrid memristive-CMOS neuromorphic processing systems}},\ }\href {https://doi.org/10.1063/1.5142089/570949} {\bibfield  {journal} {\bibinfo  {journal} {Applied Physics Letters}\ }\textbf {\bibinfo {volume} {116}},\ \bibinfo {pages} {120501} (\bibinfo {year} {2020})}\BibitemShut {NoStop}%
\bibitem [{\citenamefont {Jaeger}(2002)}]{Jaeger2002AdaptiveNetworks}%
  \BibitemOpen
  \bibfield  {author} {\bibinfo {author} {\bibfnamefont {H.}~\bibnamefont {Jaeger}},\ }\bibfield  {title} {\bibinfo {title} {{Adaptive Nonlinear System Identification with Echo State Networks}},\ }\href {http://www.ais.fraunhofer.de/INDY} {\bibfield  {journal} {\bibinfo  {journal} {Advances in Neural Information Processing Systems}\ }\textbf {\bibinfo {volume} {15}} (\bibinfo {year} {2002})}\BibitemShut {NoStop}%
\bibitem [{\citenamefont {Jaeger}\ \emph {et~al.}(2007)\citenamefont {Jaeger}, \citenamefont {Luko{\v{s}}evi{\v{c}}ius}, \citenamefont {Popovici},\ and\ \citenamefont {Siewert}}]{Jaeger2007OptimizationNeurons}%
  \BibitemOpen
  \bibfield  {author} {\bibinfo {author} {\bibfnamefont {H.}~\bibnamefont {Jaeger}}, \bibinfo {author} {\bibfnamefont {M.}~\bibnamefont {Luko{\v{s}}evi{\v{c}}ius}}, \bibinfo {author} {\bibfnamefont {D.}~\bibnamefont {Popovici}},\ and\ \bibinfo {author} {\bibfnamefont {U.}~\bibnamefont {Siewert}},\ }\bibfield  {title} {\bibinfo {title} {{Optimization and applications of echo state networks with leaky- integrator neurons}},\ }\href {https://doi.org/10.1016/J.NEUNET.2007.04.016} {\bibfield  {journal} {\bibinfo  {journal} {Neural Networks}\ }\textbf {\bibinfo {volume} {20}},\ \bibinfo {pages} {335} (\bibinfo {year} {2007})}\BibitemShut {NoStop}%
\bibitem [{\citenamefont {Jaeger}(2010)}]{Jaeger2010The1}%
  \BibitemOpen
  \bibfield  {author} {\bibinfo {author} {\bibfnamefont {H.}~\bibnamefont {Jaeger}},\ }\bibfield  {title} {\bibinfo {title} {{The "echo state" approach to analysing and training recurrent neural networks-with an Erratum note 1}},\ }\href@noop {} {\bibfield  {journal} {\bibinfo  {journal} {GMD Report}\ }\textbf {\bibinfo {volume} {148}} (\bibinfo {year} {2010})}\BibitemShut {NoStop}%
\bibitem [{\citenamefont {Yildiz}\ \emph {et~al.}(2012)\citenamefont {Yildiz}, \citenamefont {Jaeger},\ and\ \citenamefont {Kiebel}}]{Yildiz2012Re-visitingProperty}%
  \BibitemOpen
  \bibfield  {author} {\bibinfo {author} {\bibfnamefont {I.~B.}\ \bibnamefont {Yildiz}}, \bibinfo {author} {\bibfnamefont {H.}~\bibnamefont {Jaeger}},\ and\ \bibinfo {author} {\bibfnamefont {S.~J.}\ \bibnamefont {Kiebel}},\ }\bibfield  {title} {\bibinfo {title} {{Re-visiting the echo state property}},\ }\href {https://doi.org/10.1016/J.NEUNET.2012.07.005} {\bibfield  {journal} {\bibinfo  {journal} {Neural Networks}\ }\textbf {\bibinfo {volume} {35}},\ \bibinfo {pages} {1} (\bibinfo {year} {2012})}\BibitemShut {NoStop}%
\bibitem [{\citenamefont {Luko{\v{s}}evi{\v{c}}ius}(2012)}]{Lukosevicius2012ANetworks}%
  \BibitemOpen
  \bibfield  {author} {\bibinfo {author} {\bibfnamefont {M.}~\bibnamefont {Luko{\v{s}}evi{\v{c}}ius}},\ }\bibfield  {title} {\bibinfo {title} {{A Practical Guide to Applying Echo State Networks}},\ }\href {https://doi.org/10.1007/978-3-642-35289-8{\_}36} {\bibfield  {journal} {\bibinfo  {journal} {Lecture Notes in Computer Science (including subseries Lecture Notes in Artificial Intelligence and Lecture Notes in Bioinformatics)}\ }\textbf {\bibinfo {volume} {7700 LECTURE NO}},\ \bibinfo {pages} {659} (\bibinfo {year} {2012})}\BibitemShut {NoStop}%
\bibitem [{\citenamefont {Whiteaker}\ and\ \citenamefont {Gerstoft}(2021)}]{Whiteaker2021LeakySets}%
  \BibitemOpen
  \bibfield  {author} {\bibinfo {author} {\bibfnamefont {B.}~\bibnamefont {Whiteaker}}\ and\ \bibinfo {author} {\bibfnamefont {P.}~\bibnamefont {Gerstoft}},\ }\bibfield  {title} {\bibinfo {title} {{Leaky integrator dynamical systems and reachable sets}},\ }\href {https://doi.org/10.1109/ICASSP39728.2021.9413667} {\bibfield  {journal} {\bibinfo  {journal} {ICASSP, IEEE International Conference on Acoustics, Speech and Signal Processing - Proceedings}\ }\textbf {\bibinfo {volume} {2021-June}},\ \bibinfo {pages} {4025} (\bibinfo {year} {2021})}\BibitemShut {NoStop}%
\bibitem [{\citenamefont {Lin}\ \emph {et~al.}(2024)\citenamefont {Lin}, \citenamefont {Chung},\ and\ \citenamefont {Wang}}]{Lin2024AComputing}%
  \BibitemOpen
  \bibfield  {author} {\bibinfo {author} {\bibfnamefont {J.}~\bibnamefont {Lin}}, \bibinfo {author} {\bibfnamefont {F.~L.}\ \bibnamefont {Chung}},\ and\ \bibinfo {author} {\bibfnamefont {S.}~\bibnamefont {Wang}},\ }\bibfield  {title} {\bibinfo {title} {{A Fast Parametric and Structural Transfer Leaky Integrator Echo State Network for Reservoir Computing}},\ }\href {https://doi.org/10.1109/TSMC.2024.3358567} {\bibfield  {journal} {\bibinfo  {journal} {IEEE Transactions on Systems, Man, and Cybernetics: Systems}\ }\textbf {\bibinfo {volume} {54}},\ \bibinfo {pages} {3257} (\bibinfo {year} {2024})}\BibitemShut {NoStop}%
\bibitem [{\citenamefont {Sun}\ \emph {et~al.}(2024)\citenamefont {Sun}, \citenamefont {Song}, \citenamefont {Cai}, \citenamefont {Zhang}, \citenamefont {Hong},\ and\ \citenamefont {Li}}]{Sun2024AApplication}%
  \BibitemOpen
  \bibfield  {author} {\bibinfo {author} {\bibfnamefont {C.}~\bibnamefont {Sun}}, \bibinfo {author} {\bibfnamefont {M.}~\bibnamefont {Song}}, \bibinfo {author} {\bibfnamefont {D.}~\bibnamefont {Cai}}, \bibinfo {author} {\bibfnamefont {B.}~\bibnamefont {Zhang}}, \bibinfo {author} {\bibfnamefont {S.}~\bibnamefont {Hong}},\ and\ \bibinfo {author} {\bibfnamefont {H.}~\bibnamefont {Li}},\ }\bibfield  {title} {\bibinfo {title} {{A Systematic Review of Echo State Networks From Design to Application}},\ }\href {https://doi.org/10.1109/TAI.2022.3225780} {\bibfield  {journal} {\bibinfo  {journal} {IEEE Transactions on Artificial Intelligence}\ }\textbf {\bibinfo {volume} {5}},\ \bibinfo {pages} {23} (\bibinfo {year} {2024})}\BibitemShut {NoStop}%
\bibitem [{\citenamefont {Wyffels}\ \emph {et~al.}(2008)\citenamefont {Wyffels}, \citenamefont {Schrauwen}, \citenamefont {Verstraeten},\ and\ \citenamefont {Stroobandt}}]{Wyffels2008Band-passComputing}%
  \BibitemOpen
  \bibfield  {author} {\bibinfo {author} {\bibfnamefont {F.}~\bibnamefont {Wyffels}}, \bibinfo {author} {\bibfnamefont {B.}~\bibnamefont {Schrauwen}}, \bibinfo {author} {\bibfnamefont {D.}~\bibnamefont {Verstraeten}},\ and\ \bibinfo {author} {\bibfnamefont {D.}~\bibnamefont {Stroobandt}},\ }\bibfield  {title} {\bibinfo {title} {{Band-pass reservoir computing}},\ }\href {https://doi.org/10.1109/IJCNN.2008.4634252} {\bibfield  {journal} {\bibinfo  {journal} {Proceedings of the International Joint Conference on Neural Networks}\ ,\ \bibinfo {pages} {3204}} (\bibinfo {year} {2008})}\BibitemShut {NoStop}%
\bibitem [{\citenamefont {Kamsma}\ and\ \citenamefont {Teijema}(2025)}]{pyontronics}%
  \BibitemOpen
  \bibfield  {author} {\bibinfo {author} {\bibfnamefont {T.~M.}\ \bibnamefont {Kamsma}}\ and\ \bibinfo {author} {\bibfnamefont {J.~J.}\ \bibnamefont {Teijema}},\ }\href {https://doi.org/10.5281/zenodo.17076466} {\bibinfo {title} {Code repository for: Pyontronics (v0.2.0). \uppercase{Z}enodo.}} (\bibinfo {year} {2025})\BibitemShut {NoStop}%
\bibitem [{\citenamefont {Kamsma}\ \emph {et~al.}(2023{\natexlab{c}})\citenamefont {Kamsma}, \citenamefont {Boon}, \citenamefont {Spitoni},\ and\ \citenamefont {van Roij}}]{Kamsma2023UnveilingIontronics}%
  \BibitemOpen
  \bibfield  {author} {\bibinfo {author} {\bibfnamefont {T.~M.}\ \bibnamefont {Kamsma}}, \bibinfo {author} {\bibfnamefont {W.~Q.}\ \bibnamefont {Boon}}, \bibinfo {author} {\bibfnamefont {C.}~\bibnamefont {Spitoni}},\ and\ \bibinfo {author} {\bibfnamefont {R.}~\bibnamefont {van Roij}},\ }\bibfield  {title} {\bibinfo {title} {{Unveiling the capabilities of bipolar conical channels in neuromorphic iontronics}},\ }\href {https://doi.org/10.1039/D3FD00022B} {\bibfield  {journal} {\bibinfo  {journal} {Faraday Discussions}\ }\textbf {\bibinfo {volume} {246}},\ \bibinfo {pages} {125} (\bibinfo {year} {2023}{\natexlab{c}})}\BibitemShut {NoStop}%
\bibitem [{\citenamefont {Hoerl}\ and\ \citenamefont {Kennard}(1970)}]{Hoerl1970RidgeProblems}%
  \BibitemOpen
  \bibfield  {author} {\bibinfo {author} {\bibfnamefont {A.~E.}\ \bibnamefont {Hoerl}}\ and\ \bibinfo {author} {\bibfnamefont {R.~W.}\ \bibnamefont {Kennard}},\ }\bibfield  {title} {\bibinfo {title} {{Ridge Regression: Applications to Nonorthogonal Problems}},\ }\href {https://doi.org/10.1080/00401706.1970.10488635} {\bibfield  {journal} {\bibinfo  {journal} {Technometrics}\ }\textbf {\bibinfo {volume} {12}},\ \bibinfo {pages} {69} (\bibinfo {year} {1970})}\BibitemShut {NoStop}%
\bibitem [{\citenamefont {Holzmann}\ and\ \citenamefont {Hauser}(2010)}]{Holzmann2010EchoReadout}%
  \BibitemOpen
  \bibfield  {author} {\bibinfo {author} {\bibfnamefont {G.}~\bibnamefont {Holzmann}}\ and\ \bibinfo {author} {\bibfnamefont {H.}~\bibnamefont {Hauser}},\ }\bibfield  {title} {\bibinfo {title} {{Echo state networks with filter neurons and a delay{\&}sum readout}},\ }\href {https://doi.org/10.1016/J.NEUNET.2009.07.004} {\bibfield  {journal} {\bibinfo  {journal} {Neural Networks}\ }\textbf {\bibinfo {volume} {23}},\ \bibinfo {pages} {244} (\bibinfo {year} {2010})}\BibitemShut {NoStop}%
\bibitem [{\citenamefont {Schuman}\ \emph {et~al.}(2022)\citenamefont {Schuman}, \citenamefont {Kulkarni}, \citenamefont {Parsa}, \citenamefont {Mitchell}, \citenamefont {Date},\ and\ \citenamefont {Kay}}]{Schuman2022OpportunitiesApplications}%
  \BibitemOpen
  \bibfield  {author} {\bibinfo {author} {\bibfnamefont {C.~D.}\ \bibnamefont {Schuman}}, \bibinfo {author} {\bibfnamefont {S.~R.}\ \bibnamefont {Kulkarni}}, \bibinfo {author} {\bibfnamefont {M.}~\bibnamefont {Parsa}}, \bibinfo {author} {\bibfnamefont {J.~P.}\ \bibnamefont {Mitchell}}, \bibinfo {author} {\bibfnamefont {P.}~\bibnamefont {Date}},\ and\ \bibinfo {author} {\bibfnamefont {B.}~\bibnamefont {Kay}},\ }\bibfield  {title} {\bibinfo {title} {{Opportunities for neuromorphic computing algorithms and applications}},\ }\href {https://doi.org/10.1038/s43588-021-00184-y} {\bibfield  {journal} {\bibinfo  {journal} {Nature Computational Science}\ }\textbf {\bibinfo {volume} {2}},\ \bibinfo {pages} {10} (\bibinfo {year} {2022})}\BibitemShut {NoStop}%
\bibitem [{\citenamefont {Sangwan}\ and\ \citenamefont {Hersam}(2020)}]{Sangwan2020NeuromorphicMaterials}%
  \BibitemOpen
  \bibfield  {author} {\bibinfo {author} {\bibfnamefont {V.~K.}\ \bibnamefont {Sangwan}}\ and\ \bibinfo {author} {\bibfnamefont {M.~C.}\ \bibnamefont {Hersam}},\ }\bibfield  {title} {\bibinfo {title} {{Neuromorphic nanoelectronic materials}},\ }\href {https://doi.org/10.1038/s41565-020-0647-z} {\bibfield  {journal} {\bibinfo  {journal} {Nature Nanotechnology}\ }\textbf {\bibinfo {volume} {15}},\ \bibinfo {pages} {517} (\bibinfo {year} {2020})}\BibitemShut {NoStop}%
\bibitem [{\citenamefont {Zhu}\ \emph {et~al.}(2020)\citenamefont {Zhu}, \citenamefont {Zhang}, \citenamefont {Yang},\ and\ \citenamefont {Huang}}]{Zhu2020ADevices}%
  \BibitemOpen
  \bibfield  {author} {\bibinfo {author} {\bibfnamefont {J.}~\bibnamefont {Zhu}}, \bibinfo {author} {\bibfnamefont {T.}~\bibnamefont {Zhang}}, \bibinfo {author} {\bibfnamefont {Y.}~\bibnamefont {Yang}},\ and\ \bibinfo {author} {\bibfnamefont {R.}~\bibnamefont {Huang}},\ }\bibfield  {title} {\bibinfo {title} {{A comprehensive review on emerging artificial neuromorphic devices}},\ }\href {https://doi.org/10.1063/1.5118217} {\bibfield  {journal} {\bibinfo  {journal} {Applied Physics Reviews}\ }\textbf {\bibinfo {volume} {7}},\ \bibinfo {pages} {011312} (\bibinfo {year} {2020})}\BibitemShut {NoStop}%
\bibitem [{\citenamefont {Schuman}\ \emph {et~al.}(2017)\citenamefont {Schuman}, \citenamefont {Potok}, \citenamefont {Patton}, \citenamefont {Birdwell}, \citenamefont {Dean}, \citenamefont {Rose},\ and\ \citenamefont {Plank}}]{Schuman2017AHardware}%
  \BibitemOpen
  \bibfield  {author} {\bibinfo {author} {\bibfnamefont {C.~D.}\ \bibnamefont {Schuman}}, \bibinfo {author} {\bibfnamefont {T.~E.}\ \bibnamefont {Potok}}, \bibinfo {author} {\bibfnamefont {R.~M.}\ \bibnamefont {Patton}}, \bibinfo {author} {\bibfnamefont {J.~D.}\ \bibnamefont {Birdwell}}, \bibinfo {author} {\bibfnamefont {M.~E.}\ \bibnamefont {Dean}}, \bibinfo {author} {\bibfnamefont {G.~S.}\ \bibnamefont {Rose}},\ and\ \bibinfo {author} {\bibfnamefont {J.~S.}\ \bibnamefont {Plank}},\ }\bibfield  {title} {\bibinfo {title} {{A Survey of Neuromorphic Computing and Neural Networks in Hardware}},\ }\href {http://arxiv.org/abs/1705.06963} {\bibfield  {journal} {\bibinfo  {journal} {arXiv}\ } (\bibinfo {year} {2017})}\BibitemShut {NoStop}%
\bibitem [{\citenamefont {Jubin}\ \emph {et~al.}(2018)\citenamefont {Jubin}, \citenamefont {Poggioli}, \citenamefont {Siria},\ and\ \citenamefont {Bocquet}}]{Jubin2018DramaticNanopores}%
  \BibitemOpen
  \bibfield  {author} {\bibinfo {author} {\bibfnamefont {L.}~\bibnamefont {Jubin}}, \bibinfo {author} {\bibfnamefont {A.}~\bibnamefont {Poggioli}}, \bibinfo {author} {\bibfnamefont {A.}~\bibnamefont {Siria}},\ and\ \bibinfo {author} {\bibfnamefont {L.}~\bibnamefont {Bocquet}},\ }\bibfield  {title} {\bibinfo {title} {{Dramatic pressure-sensitive ion conduction in conical nanopores}},\ }\href {https://doi.org/10.1073/PNAS.1721987115/SUPPL{\_}FILE/PNAS.201721987SI.PDF} {\bibfield  {journal} {\bibinfo  {journal} {Proceedings of the National Academy of Sciences of the United States of America}\ }\textbf {\bibinfo {volume} {115}},\ \bibinfo {pages} {4063} (\bibinfo {year} {2018})}\BibitemShut {NoStop}%
\bibitem [{\citenamefont {Boon}\ \emph {et~al.}(2022)\citenamefont {Boon}, \citenamefont {Veenstra}, \citenamefont {Dijkstra},\ and\ \citenamefont {Van~Roij}}]{Boon2022Pressure-sensitiveGeometry}%
  \BibitemOpen
  \bibfield  {author} {\bibinfo {author} {\bibfnamefont {W.~Q.}\ \bibnamefont {Boon}}, \bibinfo {author} {\bibfnamefont {T.~E.}\ \bibnamefont {Veenstra}}, \bibinfo {author} {\bibfnamefont {M.}~\bibnamefont {Dijkstra}},\ and\ \bibinfo {author} {\bibfnamefont {R.}~\bibnamefont {Van~Roij}},\ }\bibfield  {title} {\bibinfo {title} {{Pressure-sensitive ion conduction in a conical channel: Optimal pressure and geometry}},\ }\href {https://doi.org/10.1063/5.0113035/2846635} {\bibfield  {journal} {\bibinfo  {journal} {Physics of Fluids}\ }\textbf {\bibinfo {volume} {34}},\ \bibinfo {pages} {101701} (\bibinfo {year} {2022})}\BibitemShut {NoStop}%
\bibitem [{\citenamefont {Barnaveli}\ \emph {et~al.}(2024)\citenamefont {Barnaveli}, \citenamefont {Kamsma}, \citenamefont {Boon},\ and\ \citenamefont {van Roij}}]{Barnaveli2024Pressure-GatedProcessing}%
  \BibitemOpen
  \bibfield  {author} {\bibinfo {author} {\bibfnamefont {A.}~\bibnamefont {Barnaveli}}, \bibinfo {author} {\bibfnamefont {T.~M.}\ \bibnamefont {Kamsma}}, \bibinfo {author} {\bibfnamefont {W.~Q.}\ \bibnamefont {Boon}},\ and\ \bibinfo {author} {\bibfnamefont {R.}~\bibnamefont {van Roij}},\ }\bibfield  {title} {\bibinfo {title} {{Pressure-Gated Microfluidic Memristor for Pulsatile Information Processing}},\ }\href {https://doi.org/10.1103/PHYSREVAPPLIED.22.054057/FIGURES/8/MEDIUM} {\bibfield  {journal} {\bibinfo  {journal} {Physical Review Applied}\ }\textbf {\bibinfo {volume} {22}},\ \bibinfo {pages} {054057} (\bibinfo {year} {2024})}\BibitemShut {NoStop}%
\bibitem [{\citenamefont {Van Der~Heyden}\ \emph {et~al.}(2005)\citenamefont {Van Der~Heyden}, \citenamefont {Stein},\ and\ \citenamefont {Dekker}}]{VanDerHeyden2005StreamingChannel}%
  \BibitemOpen
  \bibfield  {author} {\bibinfo {author} {\bibfnamefont {F.~H.~J.}\ \bibnamefont {Van Der~Heyden}}, \bibinfo {author} {\bibfnamefont {D.}~\bibnamefont {Stein}},\ and\ \bibinfo {author} {\bibfnamefont {C.}~\bibnamefont {Dekker}},\ }\bibfield  {title} {\bibinfo {title} {{Streaming Currents in a Single Nanofluidic Channel}},\ }\href {https://doi.org/10.1103/PhysRevLett.95.116104} {\bibfield  {journal} {\bibinfo  {journal} {Physical Review Letters}\ }\textbf {\bibinfo {volume} {95}},\ \bibinfo {pages} {116104} (\bibinfo {year} {2005})}\BibitemShut {NoStop}%
\bibitem [{\citenamefont {Werkhoven}\ and\ \citenamefont {Van~Roij}(2020)}]{Werkhoven2020CoupledSystems}%
  \BibitemOpen
  \bibfield  {author} {\bibinfo {author} {\bibfnamefont {B.~L.}\ \bibnamefont {Werkhoven}}\ and\ \bibinfo {author} {\bibfnamefont {R.}~\bibnamefont {Van~Roij}},\ }\bibfield  {title} {\bibinfo {title} {{Coupled water, charge and salt transport in heterogeneous nano-fluidic systems}},\ }\href {https://doi.org/10.1039/C9SM02144B} {\bibfield  {journal} {\bibinfo  {journal} {Soft Matter}\ }\textbf {\bibinfo {volume} {16}},\ \bibinfo {pages} {1527} (\bibinfo {year} {2020})}\BibitemShut {NoStop}%
\bibitem [{\citenamefont {Kamsma}\ \emph {et~al.}(2024{\natexlab{b}})\citenamefont {Kamsma}, \citenamefont {van Roij},\ and\ \citenamefont {Spitoni}}]{Kamsma2024ACircuits}%
  \BibitemOpen
  \bibfield  {author} {\bibinfo {author} {\bibfnamefont {T.}~\bibnamefont {Kamsma}}, \bibinfo {author} {\bibfnamefont {R.}~\bibnamefont {van Roij}},\ and\ \bibinfo {author} {\bibfnamefont {C.}~\bibnamefont {Spitoni}},\ }\bibfield  {title} {\bibinfo {title} {{A simple mathematical theory for Simple Volatile Memristors and their spiking circuits}},\ }\href {https://doi.org/10.1016/J.CHAOS.2024.115320} {\bibfield  {journal} {\bibinfo  {journal} {Chaos, Solitons {\&} Fractals}\ }\textbf {\bibinfo {volume} {186}},\ \bibinfo {pages} {115320} (\bibinfo {year} {2024}{\natexlab{b}})}\BibitemShut {NoStop}%
\bibitem [{\citenamefont {Shi}\ \emph {et~al.}(2023)\citenamefont {Shi}, \citenamefont {Wang}, \citenamefont {Liang}, \citenamefont {Duan}, \citenamefont {Du},\ and\ \citenamefont {Xie}}]{Shi2023UltralowMemristor}%
  \BibitemOpen
  \bibfield  {author} {\bibinfo {author} {\bibfnamefont {D.}~\bibnamefont {Shi}}, \bibinfo {author} {\bibfnamefont {W.}~\bibnamefont {Wang}}, \bibinfo {author} {\bibfnamefont {Y.}~\bibnamefont {Liang}}, \bibinfo {author} {\bibfnamefont {L.}~\bibnamefont {Duan}}, \bibinfo {author} {\bibfnamefont {G.}~\bibnamefont {Du}},\ and\ \bibinfo {author} {\bibfnamefont {Y.}~\bibnamefont {Xie}},\ }\bibfield  {title} {\bibinfo {title} {{Ultralow Energy Consumption Angstrom-Fluidic Memristor}},\ }\bibfield  {journal} {\bibinfo  {journal} {Nano Letters}\ }\href {https://doi.org/10.1021/ACS.NANOLETT.3C03518} {10.1021/ACS.NANOLETT.3C03518} (\bibinfo {year} {2023})\BibitemShut {NoStop}%
\bibitem [{\citenamefont {Storm}\ \emph {et~al.}(2003)\citenamefont {Storm}, \citenamefont {Chen}, \citenamefont {Ling}, \citenamefont {Zandbergen},\ and\ \citenamefont {Dekker}}]{Storm2003FabricationPrecision}%
  \BibitemOpen
  \bibfield  {author} {\bibinfo {author} {\bibfnamefont {A.~J.}\ \bibnamefont {Storm}}, \bibinfo {author} {\bibfnamefont {J.~H.}\ \bibnamefont {Chen}}, \bibinfo {author} {\bibfnamefont {X.~S.}\ \bibnamefont {Ling}}, \bibinfo {author} {\bibfnamefont {H.~W.}\ \bibnamefont {Zandbergen}},\ and\ \bibinfo {author} {\bibfnamefont {C.}~\bibnamefont {Dekker}},\ }\bibfield  {title} {\bibinfo {title} {{Fabrication of solid-state nanopores with single-nanometre precision}},\ }\href {https://doi.org/10.1038/NMAT941} {\bibfield  {journal} {\bibinfo  {journal} {Nature Materials 2003 2:8}\ }\textbf {\bibinfo {volume} {2}},\ \bibinfo {pages} {537} (\bibinfo {year} {2003})}\BibitemShut {NoStop}%
\bibitem [{\citenamefont {Yang}\ \emph {et~al.}(2016)\citenamefont {Yang}, \citenamefont {Hinkle}, \citenamefont {Menestrina}, \citenamefont {Vlassiouk},\ and\ \citenamefont {Siwy}}]{Yang2016PolarizationRectification}%
  \BibitemOpen
  \bibfield  {author} {\bibinfo {author} {\bibfnamefont {C.}~\bibnamefont {Yang}}, \bibinfo {author} {\bibfnamefont {P.}~\bibnamefont {Hinkle}}, \bibinfo {author} {\bibfnamefont {J.}~\bibnamefont {Menestrina}}, \bibinfo {author} {\bibfnamefont {I.~V.}\ \bibnamefont {Vlassiouk}},\ and\ \bibinfo {author} {\bibfnamefont {Z.~S.}\ \bibnamefont {Siwy}},\ }\bibfield  {title} {\bibinfo {title} {{Polarization of Gold in Nanopores Leads to Ion Current Rectification}},\ }\href {https://doi.org/10.1021/ACS.JPCLETT.6B01971/ASSET/IMAGES/LARGE/JZ-2016-01971F{\_}0003.JPEG} {\bibfield  {journal} {\bibinfo  {journal} {Journal of Physical Chemistry Letters}\ }\textbf {\bibinfo {volume} {7}},\ \bibinfo {pages} {4152} (\bibinfo {year} {2016})}\BibitemShut {NoStop}%
\bibitem [{\citenamefont {Zhou}\ \emph {et~al.}(2023)\citenamefont {Zhou}, \citenamefont {Eden}, \citenamefont {Chou}, \citenamefont {Huber},\ and\ \citenamefont {Pennathur}}]{Zhou2023NanofluidicNanochannels}%
  \BibitemOpen
  \bibfield  {author} {\bibinfo {author} {\bibfnamefont {L.}~\bibnamefont {Zhou}}, \bibinfo {author} {\bibfnamefont {A.}~\bibnamefont {Eden}}, \bibinfo {author} {\bibfnamefont {K.~H.}\ \bibnamefont {Chou}}, \bibinfo {author} {\bibfnamefont {D.~E.}\ \bibnamefont {Huber}},\ and\ \bibinfo {author} {\bibfnamefont {S.}~\bibnamefont {Pennathur}},\ }\bibfield  {title} {\bibinfo {title} {{Nanofluidic diodes based on asymmetric bio-inspired surface coatings in straight glass nanochannels}},\ }\href {https://doi.org/10.1039/D3FD00074E} {\bibfield  {journal} {\bibinfo  {journal} {Faraday Discussions}\ }\textbf {\bibinfo {volume} {246}},\ \bibinfo {pages} {356} (\bibinfo {year} {2023})}\BibitemShut {NoStop}%
\bibitem [{\citenamefont {Wang}\ \emph {et~al.}(2012)\citenamefont {Wang}, \citenamefont {Kvetny}, \citenamefont {Liu}, \citenamefont {Brown}, \citenamefont {Li},\ and\ \citenamefont {Wang}}]{Wang2012TransmembraneTransport}%
  \BibitemOpen
  \bibfield  {author} {\bibinfo {author} {\bibfnamefont {D.}~\bibnamefont {Wang}}, \bibinfo {author} {\bibfnamefont {M.}~\bibnamefont {Kvetny}}, \bibinfo {author} {\bibfnamefont {J.}~\bibnamefont {Liu}}, \bibinfo {author} {\bibfnamefont {W.}~\bibnamefont {Brown}}, \bibinfo {author} {\bibfnamefont {Y.}~\bibnamefont {Li}},\ and\ \bibinfo {author} {\bibfnamefont {G.}~\bibnamefont {Wang}},\ }\bibfield  {title} {\bibinfo {title} {{Transmembrane potential across single conical nanopores and resulting memristive and memcapacitive ion transport}},\ }\href {https://doi.org/10.1021/ja211142e} {\bibfield  {journal} {\bibinfo  {journal} {Journal of the American Chemical Society}\ }\textbf {\bibinfo {volume} {134}},\ \bibinfo {pages} {3651} (\bibinfo {year} {2012})}\BibitemShut {NoStop}%
\bibitem [{\citenamefont {Li}\ \emph {et~al.}(2018)\citenamefont {Li}, \citenamefont {Belkin}, \citenamefont {Li}, \citenamefont {Yan}, \citenamefont {Hu}, \citenamefont {Ge}, \citenamefont {Jiang}, \citenamefont {Montgomery}, \citenamefont {Lin}, \citenamefont {Wang}, \citenamefont {Song}, \citenamefont {Strachan}, \citenamefont {Barnell}, \citenamefont {Wu}, \citenamefont {Williams}, \citenamefont {Yang},\ and\ \citenamefont {Xia}}]{Li2018EfficientNetworks}%
  \BibitemOpen
  \bibfield  {author} {\bibinfo {author} {\bibfnamefont {C.}~\bibnamefont {Li}}, \bibinfo {author} {\bibfnamefont {D.}~\bibnamefont {Belkin}}, \bibinfo {author} {\bibfnamefont {Y.}~\bibnamefont {Li}}, \bibinfo {author} {\bibfnamefont {P.}~\bibnamefont {Yan}}, \bibinfo {author} {\bibfnamefont {M.}~\bibnamefont {Hu}}, \bibinfo {author} {\bibfnamefont {N.}~\bibnamefont {Ge}}, \bibinfo {author} {\bibfnamefont {H.}~\bibnamefont {Jiang}}, \bibinfo {author} {\bibfnamefont {E.}~\bibnamefont {Montgomery}}, \bibinfo {author} {\bibfnamefont {P.}~\bibnamefont {Lin}}, \bibinfo {author} {\bibfnamefont {Z.}~\bibnamefont {Wang}}, \bibinfo {author} {\bibfnamefont {W.}~\bibnamefont {Song}}, \bibinfo {author} {\bibfnamefont {J.~P.}\ \bibnamefont {Strachan}}, \bibinfo {author} {\bibfnamefont {M.}~\bibnamefont {Barnell}}, \bibinfo {author} {\bibfnamefont {Q.}~\bibnamefont {Wu}}, \bibinfo {author} {\bibfnamefont {R.~S.}\ \bibnamefont {Williams}}, \bibinfo {author} {\bibfnamefont {J.~J.}\ \bibnamefont {Yang}},\ and\ \bibinfo
  {author} {\bibfnamefont {Q.}~\bibnamefont {Xia}},\ }\bibfield  {title} {\bibinfo {title} {{Efficient and self-adaptive in-situ learning in multilayer memristor neural networks}},\ }\href {https://doi.org/10.1038/s41467-018-04484-2} {\bibfield  {journal} {\bibinfo  {journal} {Nature Communications 2018 9:1}\ }\textbf {\bibinfo {volume} {9}},\ \bibinfo {pages} {1} (\bibinfo {year} {2018})}\BibitemShut {NoStop}%
\bibitem [{\citenamefont {Liu}\ \emph {et~al.}(2025)\citenamefont {Liu}, \citenamefont {Wang}, \citenamefont {Sun}, \citenamefont {Lu}, \citenamefont {Shi},\ and\ \citenamefont {Xie}}]{Liu2025Resistance-RestorableChip}%
  \BibitemOpen
  \bibfield  {author} {\bibinfo {author} {\bibfnamefont {K.}~\bibnamefont {Liu}}, \bibinfo {author} {\bibfnamefont {Y.}~\bibnamefont {Wang}}, \bibinfo {author} {\bibfnamefont {M.}~\bibnamefont {Sun}}, \bibinfo {author} {\bibfnamefont {J.}~\bibnamefont {Lu}}, \bibinfo {author} {\bibfnamefont {D.}~\bibnamefont {Shi}},\ and\ \bibinfo {author} {\bibfnamefont {Y.}~\bibnamefont {Xie}},\ }\bibfield  {title} {\bibinfo {title} {{Resistance-Restorable Nanofluidic Memristor and Neuromorphic Chip}},\ }\bibfield  {journal} {\bibinfo  {journal} {Nano Letters}\ }\href {https://doi.org/10.1021/ACS.NANOLETT.5C00315} {10.1021/ACS.NANOLETT.5C00315} (\bibinfo {year} {2025})\BibitemShut {NoStop}%
\bibitem [{\citenamefont {Van De~Burgt}\ \emph {et~al.}(2018)\citenamefont {Van De~Burgt}, \citenamefont {Melianas}, \citenamefont {Keene}, \citenamefont {Malliaras},\ and\ \citenamefont {Salleo}}]{VanDeBurgt2018OrganicComputing}%
  \BibitemOpen
  \bibfield  {author} {\bibinfo {author} {\bibfnamefont {Y.}~\bibnamefont {Van De~Burgt}}, \bibinfo {author} {\bibfnamefont {A.}~\bibnamefont {Melianas}}, \bibinfo {author} {\bibfnamefont {S.~T.}\ \bibnamefont {Keene}}, \bibinfo {author} {\bibfnamefont {G.}~\bibnamefont {Malliaras}},\ and\ \bibinfo {author} {\bibfnamefont {A.}~\bibnamefont {Salleo}},\ }\bibfield  {title} {\bibinfo {title} {{Organic electronics for neuromorphic computing}},\ }\href {https://doi.org/10.1038/s41928-018-0103-3} {\bibfield  {journal} {\bibinfo  {journal} {Nature Electronics}\ }\textbf {\bibinfo {volume} {1}},\ \bibinfo {pages} {386} (\bibinfo {year} {2018})}\BibitemShut {NoStop}%
\bibitem [{\citenamefont {Kazemzadeh}\ \emph {et~al.}(2025)\citenamefont {Kazemzadeh}, \citenamefont {Stevens},\ and\ \citenamefont {van~de Burgt}}]{Kazemzadeh2025AllArray}%
  \BibitemOpen
  \bibfield  {author} {\bibinfo {author} {\bibfnamefont {S.}~\bibnamefont {Kazemzadeh}}, \bibinfo {author} {\bibfnamefont {T.}~\bibnamefont {Stevens}},\ and\ \bibinfo {author} {\bibfnamefont {Y.}~\bibnamefont {van~de Burgt}},\ }\bibfield  {title} {\bibinfo {title} {{All Organic Fully Integrated Neuromorphic Crossbar Array}},\ }\href {https://doi.org/10.1002/AELM.202500054;CSUBTYPE:STRING:AHEAD} {\bibfield  {journal} {\bibinfo  {journal} {Advanced Electronic Materials}\ ,\ \bibinfo {pages} {2500054}} (\bibinfo {year} {2025})}\BibitemShut {NoStop}%
\bibitem [{\citenamefont {Xu}\ \emph {et~al.}(2025)\citenamefont {Xu}, \citenamefont {Cui}, \citenamefont {Wang}, \citenamefont {Zhang}, \citenamefont {Liu}, \citenamefont {Mei}, \citenamefont {Wu}, \citenamefont {Wan},\ and\ \citenamefont {Xiao}}]{Xu2025Angstrom-Scale-ChannelComputing}%
  \BibitemOpen
  \bibfield  {author} {\bibinfo {author} {\bibfnamefont {G.}~\bibnamefont {Xu}}, \bibinfo {author} {\bibfnamefont {H.}~\bibnamefont {Cui}}, \bibinfo {author} {\bibfnamefont {L.}~\bibnamefont {Wang}}, \bibinfo {author} {\bibfnamefont {M.}~\bibnamefont {Zhang}}, \bibinfo {author} {\bibfnamefont {W.}~\bibnamefont {Liu}}, \bibinfo {author} {\bibfnamefont {T.}~\bibnamefont {Mei}}, \bibinfo {author} {\bibfnamefont {B.}~\bibnamefont {Wu}}, \bibinfo {author} {\bibfnamefont {C.}~\bibnamefont {Wan}},\ and\ \bibinfo {author} {\bibfnamefont {K.}~\bibnamefont {Xiao}},\ }\bibfield  {title} {\bibinfo {title} {{{\AA}ngstr{\"{o}}m-Scale-Channel Iontronic Memristors for Neuromorphic Computing}},\ }\href {https://doi.org/10.1021/ACSAMI.5C01409/ASSET/IMAGES/LARGE/AM5C01409{\_}0005.JPEG} {\bibfield  {journal} {\bibinfo  {journal} {ACS Applied Materials and Interfaces}\ }\textbf {\bibinfo {volume} {17}},\ \bibinfo {pages} {34659} (\bibinfo {year} {2025})}\BibitemShut {NoStop}%
\bibitem [{\citenamefont {Zhang}\ \emph {et~al.}(2022)\citenamefont {Zhang}, \citenamefont {van Doremaele}, \citenamefont {Ye}, \citenamefont {Stevens}, \citenamefont {Song}, \citenamefont {Chiechi},\ and\ \citenamefont {van~de Burgt}}]{Zhang2022AdaptiveConductor}%
  \BibitemOpen
  \bibfield  {author} {\bibinfo {author} {\bibfnamefont {Y.}~\bibnamefont {Zhang}}, \bibinfo {author} {\bibfnamefont {E.~R.~W.}\ \bibnamefont {van Doremaele}}, \bibinfo {author} {\bibfnamefont {G.}~\bibnamefont {Ye}}, \bibinfo {author} {\bibfnamefont {T.}~\bibnamefont {Stevens}}, \bibinfo {author} {\bibfnamefont {J.}~\bibnamefont {Song}}, \bibinfo {author} {\bibfnamefont {R.~C.}\ \bibnamefont {Chiechi}},\ and\ \bibinfo {author} {\bibfnamefont {Y.}~\bibnamefont {van~de Burgt}},\ }\bibfield  {title} {\bibinfo {title} {{Adaptive Biosensing and Neuromorphic Classification Based on an Ambipolar Organic Mixed Ionic-Electronic Conductor}},\ }\href {https://doi.org/10.1002/adma.202200393} {\bibfield  {journal} {\bibinfo  {journal} {Advanced Materials}\ ,\ \bibinfo {pages} {2200393}} (\bibinfo {year} {2022})}\BibitemShut {NoStop}%
\bibitem [{\citenamefont {Hu}\ \emph {et~al.}(2023)\citenamefont {Hu}, \citenamefont {Fattori}, \citenamefont {Schilp}, \citenamefont {Verbeek}, \citenamefont {Kazemzadeh}, \citenamefont {van~de Burgt}, \citenamefont {Kronemeijer}, \citenamefont {Gelinck},\ and\ \citenamefont {Cantatore}}]{Hu2023AnComputing}%
  \BibitemOpen
  \bibfield  {author} {\bibinfo {author} {\bibfnamefont {L.~S.}\ \bibnamefont {Hu}}, \bibinfo {author} {\bibfnamefont {M.}~\bibnamefont {Fattori}}, \bibinfo {author} {\bibfnamefont {W.}~\bibnamefont {Schilp}}, \bibinfo {author} {\bibfnamefont {R.}~\bibnamefont {Verbeek}}, \bibinfo {author} {\bibfnamefont {S.}~\bibnamefont {Kazemzadeh}}, \bibinfo {author} {\bibfnamefont {Y.}~\bibnamefont {van~de Burgt}}, \bibinfo {author} {\bibfnamefont {A.~J.}\ \bibnamefont {Kronemeijer}}, \bibinfo {author} {\bibfnamefont {G.}~\bibnamefont {Gelinck}},\ and\ \bibinfo {author} {\bibfnamefont {E.}~\bibnamefont {Cantatore}},\ }\bibfield  {title} {\bibinfo {title} {{An Energy-Efficient Solid-State Organic Device Array for Neuromorphic Computing}},\ }\href {https://doi.org/10.1109/TED.2023.3327947} {\bibfield  {journal} {\bibinfo  {journal} {IEEE Transactions on Electron Devices}\ }\textbf {\bibinfo {volume} {70}},\ \bibinfo {pages} {6520} (\bibinfo {year} {2023})}\BibitemShut {NoStop}%
\bibitem [{\citenamefont {Van De~Burgt}\ \emph {et~al.}(2017)\citenamefont {Van De~Burgt}, \citenamefont {Lubberman}, \citenamefont {Fuller}, \citenamefont {Keene}, \citenamefont {Faria}, \citenamefont {Agarwal}, \citenamefont {Marinella}, \citenamefont {Alec~Talin},\ and\ \citenamefont {Salleo}}]{VanDeBurgt2017AComputing}%
  \BibitemOpen
  \bibfield  {author} {\bibinfo {author} {\bibfnamefont {Y.}~\bibnamefont {Van De~Burgt}}, \bibinfo {author} {\bibfnamefont {E.}~\bibnamefont {Lubberman}}, \bibinfo {author} {\bibfnamefont {E.~J.}\ \bibnamefont {Fuller}}, \bibinfo {author} {\bibfnamefont {S.~T.}\ \bibnamefont {Keene}}, \bibinfo {author} {\bibfnamefont {G.~C.}\ \bibnamefont {Faria}}, \bibinfo {author} {\bibfnamefont {S.}~\bibnamefont {Agarwal}}, \bibinfo {author} {\bibfnamefont {M.~J.}\ \bibnamefont {Marinella}}, \bibinfo {author} {\bibfnamefont {A.}~\bibnamefont {Alec~Talin}},\ and\ \bibinfo {author} {\bibfnamefont {A.}~\bibnamefont {Salleo}},\ }\bibfield  {title} {\bibinfo {title} {{A non-volatile organic electrochemical device as a low-voltage artificial synapse for neuromorphic computing}},\ }\href {https://doi.org/10.1038/NMAT4856;TECHMETA=120,128,140,146;SUBJMETA=1005,301,639,923;KWRD=MATERIALS+FOR+DEVICES,SOFT+MATERIALS} {\bibfield  {journal} {\bibinfo  {journal} {Nature Materials}\ }\textbf {\bibinfo {volume} {16}},\ \bibinfo {pages}
  {414} (\bibinfo {year} {2017})}\BibitemShut {NoStop}%
\bibitem [{\citenamefont {L{\'{o}}pez-Caraballo}\ \emph {et~al.}(2016)\citenamefont {L{\'{o}}pez-Caraballo}, \citenamefont {Salfate}, \citenamefont {Lazz{\'{u}}s}, \citenamefont {Rojas}, \citenamefont {Rivera},\ and\ \citenamefont {Palma-Chilla}}]{Lopez-Caraballo2016Mackey-GlassNetwork}%
  \BibitemOpen
  \bibfield  {author} {\bibinfo {author} {\bibfnamefont {C.~H.}\ \bibnamefont {L{\'{o}}pez-Caraballo}}, \bibinfo {author} {\bibfnamefont {I.}~\bibnamefont {Salfate}}, \bibinfo {author} {\bibfnamefont {J.~A.}\ \bibnamefont {Lazz{\'{u}}s}}, \bibinfo {author} {\bibfnamefont {P.}~\bibnamefont {Rojas}}, \bibinfo {author} {\bibfnamefont {M.}~\bibnamefont {Rivera}},\ and\ \bibinfo {author} {\bibfnamefont {L.}~\bibnamefont {Palma-Chilla}},\ }\bibfield  {title} {\bibinfo {title} {{Mackey-Glass noisy chaotic time series prediction by a swarm-optimized neural network}},\ }\href {https://doi.org/10.1088/1742-6596/720/1/012002} {\bibfield  {journal} {\bibinfo  {journal} {Journal of Physics: Conference Series}\ }\textbf {\bibinfo {volume} {720}},\ \bibinfo {pages} {012002} (\bibinfo {year} {2016})}\BibitemShut {NoStop}%
\bibitem [{\citenamefont {Chen}\ \emph {et~al.}(2006)\citenamefont {Chen}, \citenamefont {Yang},\ and\ \citenamefont {Dong}}]{Chen2006Time-seriesNetwork}%
  \BibitemOpen
  \bibfield  {author} {\bibinfo {author} {\bibfnamefont {Y.}~\bibnamefont {Chen}}, \bibinfo {author} {\bibfnamefont {B.}~\bibnamefont {Yang}},\ and\ \bibinfo {author} {\bibfnamefont {J.}~\bibnamefont {Dong}},\ }\bibfield  {title} {\bibinfo {title} {{Time-series prediction using a local linear wavelet neural network}},\ }\href {https://doi.org/10.1016/J.NEUCOM.2005.02.006} {\bibfield  {journal} {\bibinfo  {journal} {Neurocomputing}\ }\textbf {\bibinfo {volume} {69}},\ \bibinfo {pages} {449} (\bibinfo {year} {2006})}\BibitemShut {NoStop}%
\bibitem [{\citenamefont {Mackey}\ and\ \citenamefont {Glass}(1977)}]{Mackey1977OscillationSystems}%
  \BibitemOpen
  \bibfield  {author} {\bibinfo {author} {\bibfnamefont {M.~C.}\ \bibnamefont {Mackey}}\ and\ \bibinfo {author} {\bibfnamefont {L.}~\bibnamefont {Glass}},\ }\bibfield  {title} {\bibinfo {title} {{Oscillation and Chaos in Physiological Control Systems}},\ }\href {https://doi.org/10.1126/SCIENCE.267326} {\bibfield  {journal} {\bibinfo  {journal} {Science}\ }\textbf {\bibinfo {volume} {197}},\ \bibinfo {pages} {287} (\bibinfo {year} {1977})}\BibitemShut {NoStop}%
\bibitem [{\citenamefont {Soltani}\ \emph {et~al.}(2023)\citenamefont {Soltani}, \citenamefont {Benmohamed},\ and\ \citenamefont {Ltifi}}]{Soltani2023EchoReview}%
  \BibitemOpen
  \bibfield  {author} {\bibinfo {author} {\bibfnamefont {R.}~\bibnamefont {Soltani}}, \bibinfo {author} {\bibfnamefont {E.}~\bibnamefont {Benmohamed}},\ and\ \bibinfo {author} {\bibfnamefont {H.}~\bibnamefont {Ltifi}},\ }\bibfield  {title} {\bibinfo {title} {{Echo State Network Optimization: A Systematic Literature Review}},\ }\href {https://doi.org/10.1007/S11063-023-11326-W/FIGURES/9} {\bibfield  {journal} {\bibinfo  {journal} {Neural Processing Letters}\ }\textbf {\bibinfo {volume} {55}},\ \bibinfo {pages} {10251} (\bibinfo {year} {2023})}\BibitemShut {NoStop}%
\bibitem [{\citenamefont {Doyne~Farmer}(1982)}]{DoyneFarmer1982ChaoticSystem}%
  \BibitemOpen
  \bibfield  {author} {\bibinfo {author} {\bibfnamefont {J.}~\bibnamefont {Doyne~Farmer}},\ }\bibfield  {title} {\bibinfo {title} {{Chaotic attractors of an infinite-dimensional dynamical system}},\ }\href {https://doi.org/10.1016/0167-2789(82)90042-2} {\bibfield  {journal} {\bibinfo  {journal} {Physica D: Nonlinear Phenomena}\ }\textbf {\bibinfo {volume} {4}},\ \bibinfo {pages} {366} (\bibinfo {year} {1982})}\BibitemShut {NoStop}%
\bibitem [{\citenamefont {Vesanto}(1997)}]{Vesanto1997UsingPrediction}%
  \BibitemOpen
  \bibfield  {author} {\bibinfo {author} {\bibfnamefont {J.}~\bibnamefont {Vesanto}},\ }\bibfield  {title} {\bibinfo {title} {{Using the SOM and Local Models in Time-Series Prediction Using the SOM and Local Models in Time-Series Prediction}},\ }\href@noop {} {\bibfield  {journal} {\bibinfo  {journal} {Proceedings of Workshop on Self-Organizing Maps}\ ,\ \bibinfo {pages} {209}} (\bibinfo {year} {1997})}\BibitemShut {NoStop}%
\bibitem [{\citenamefont {Gers}\ \emph {et~al.}(2001)\citenamefont {Gers}, \citenamefont {Eck},\ and\ \citenamefont {Schmidhuber}}]{Gers2001ApplyingApproaches}%
  \BibitemOpen
  \bibfield  {author} {\bibinfo {author} {\bibfnamefont {F.~A.}\ \bibnamefont {Gers}}, \bibinfo {author} {\bibfnamefont {D.}~\bibnamefont {Eck}},\ and\ \bibinfo {author} {\bibfnamefont {J.}~\bibnamefont {Schmidhuber}},\ }\bibfield  {title} {\bibinfo {title} {{Applying LSTM to Time Series Predictable through Time-Window Approaches}},\ }\href {https://doi.org/10.1007/3-540-44668-0{\_}93} {\bibfield  {journal} {\bibinfo  {journal} {Lecture Notes in Computer Science (including subseries Lecture Notes in Artificial Intelligence and Lecture Notes in Bioinformatics)}\ }\textbf {\bibinfo {volume} {2130}},\ \bibinfo {pages} {669} (\bibinfo {year} {2001})}\BibitemShut {NoStop}%
\bibitem [{\citenamefont {Ustundag}\ and\ \citenamefont {Kulaglic}(2020)}]{Ustundag2020High-PerformanceNetworks}%
  \BibitemOpen
  \bibfield  {author} {\bibinfo {author} {\bibfnamefont {B.~B.}\ \bibnamefont {Ustundag}}\ and\ \bibinfo {author} {\bibfnamefont {A.}~\bibnamefont {Kulaglic}},\ }\bibfield  {title} {\bibinfo {title} {{High-Performance Time Series Prediction with Predictive Error Compensated Wavelet Neural Networks}},\ }\href {https://doi.org/10.1109/ACCESS.2020.3038724} {\bibfield  {journal} {\bibinfo  {journal} {IEEE Access}\ }\textbf {\bibinfo {volume} {8}},\ \bibinfo {pages} {210532} (\bibinfo {year} {2020})}\BibitemShut {NoStop}%
\bibitem [{\citenamefont {Akiba}\ \emph {et~al.}(2019)\citenamefont {Akiba}, \citenamefont {Sano}, \citenamefont {Yanase}, \citenamefont {Ohta},\ and\ \citenamefont {Koyama}}]{Akiba2019Optuna:Framework}%
  \BibitemOpen
  \bibfield  {author} {\bibinfo {author} {\bibfnamefont {T.}~\bibnamefont {Akiba}}, \bibinfo {author} {\bibfnamefont {S.}~\bibnamefont {Sano}}, \bibinfo {author} {\bibfnamefont {T.}~\bibnamefont {Yanase}}, \bibinfo {author} {\bibfnamefont {T.}~\bibnamefont {Ohta}},\ and\ \bibinfo {author} {\bibfnamefont {M.}~\bibnamefont {Koyama}},\ }\bibfield  {title} {\bibinfo {title} {{Optuna: A Next-generation Hyperparameter Optimization Framework}},\ }\href {https://doi.org/10.1145/3292500.3330701} {\bibfield  {journal} {\bibinfo  {journal} {Proceedings of the ACM SIGKDD International Conference on Knowledge Discovery and Data Mining}\ ,\ \bibinfo {pages} {2623}} (\bibinfo {year} {2019})}\BibitemShut {NoStop}%
\bibitem [{\citenamefont {Howard}\ \emph {et~al.}(2021)\citenamefont {Howard}, \citenamefont {alexjyu}, \citenamefont {Suo},\ and\ \citenamefont {Cukierski}}]{ventilator-pressure-prediction}%
  \BibitemOpen
  \bibfield  {author} {\bibinfo {author} {\bibfnamefont {A.}~\bibnamefont {Howard}}, \bibinfo {author} {\bibnamefont {alexjyu}}, \bibinfo {author} {\bibfnamefont {D.}~\bibnamefont {Suo}},\ and\ \bibinfo {author} {\bibfnamefont {W.}~\bibnamefont {Cukierski}},\ }\href@noop {} {\bibinfo {title} {Google brain - ventilator pressure prediction}},\ \bibinfo {howpublished} {\url{https://kaggle.com/competitions/ventilator-pressure-prediction}} (\bibinfo {year} {2021}),\ \bibinfo {note} {kaggle}\BibitemShut {NoStop}%
\bibitem [{\citenamefont {Xiong}\ \emph {et~al.}(2023)\citenamefont {Xiong}, \citenamefont {Li}, \citenamefont {He}, \citenamefont {Xie}, \citenamefont {Zong}, \citenamefont {Jiang}, \citenamefont {Ma}, \citenamefont {Wu}, \citenamefont {Fei}, \citenamefont {Yu},\ and\ \citenamefont {Mao}}]{Xiong2023NeuromorphicMemristor}%
  \BibitemOpen
  \bibfield  {author} {\bibinfo {author} {\bibfnamefont {T.}~\bibnamefont {Xiong}}, \bibinfo {author} {\bibfnamefont {C.}~\bibnamefont {Li}}, \bibinfo {author} {\bibfnamefont {X.}~\bibnamefont {He}}, \bibinfo {author} {\bibfnamefont {B.}~\bibnamefont {Xie}}, \bibinfo {author} {\bibfnamefont {J.}~\bibnamefont {Zong}}, \bibinfo {author} {\bibfnamefont {Y.}~\bibnamefont {Jiang}}, \bibinfo {author} {\bibfnamefont {W.}~\bibnamefont {Ma}}, \bibinfo {author} {\bibfnamefont {F.}~\bibnamefont {Wu}}, \bibinfo {author} {\bibfnamefont {J.}~\bibnamefont {Fei}}, \bibinfo {author} {\bibfnamefont {P.}~\bibnamefont {Yu}},\ and\ \bibinfo {author} {\bibfnamefont {L.}~\bibnamefont {Mao}},\ }\bibfield  {title} {\bibinfo {title} {{Neuromorphic functions with a polyelectrolyte-confined fluidic memristor}},\ }\href {https://doi.org/10.1126/science.adc9150} {\bibfield  {journal} {\bibinfo  {journal} {Science}\ }\textbf {\bibinfo {volume} {379}},\ \bibinfo {pages} {156} (\bibinfo {year} {2023})}\BibitemShut {NoStop}%
\bibitem [{\citenamefont {Ling}\ \emph {et~al.}(2024)\citenamefont {Ling}, \citenamefont {Yu}, \citenamefont {Guo}, \citenamefont {Bian}, \citenamefont {Wang}, \citenamefont {Wang}, \citenamefont {Hou},\ and\ \citenamefont {Hou}}]{Ling2024Single-PoreCombinations}%
  \BibitemOpen
  \bibfield  {author} {\bibinfo {author} {\bibfnamefont {Y.}~\bibnamefont {Ling}}, \bibinfo {author} {\bibfnamefont {L.}~\bibnamefont {Yu}}, \bibinfo {author} {\bibfnamefont {Z.}~\bibnamefont {Guo}}, \bibinfo {author} {\bibfnamefont {F.}~\bibnamefont {Bian}}, \bibinfo {author} {\bibfnamefont {Y.}~\bibnamefont {Wang}}, \bibinfo {author} {\bibfnamefont {X.}~\bibnamefont {Wang}}, \bibinfo {author} {\bibfnamefont {Y.}~\bibnamefont {Hou}},\ and\ \bibinfo {author} {\bibfnamefont {X.}~\bibnamefont {Hou}},\ }\bibfield  {title} {\bibinfo {title} {{Single-Pore Nanofluidic Logic Memristor with Reconfigurable Synaptic Functions and Designable Combinations}},\ }\bibfield  {journal} {\bibinfo  {journal} {Journal of the American Chemical Society}\ }\href {https://doi.org/10.1021/JACS.4C01218} {10.1021/JACS.4C01218} (\bibinfo {year} {2024})\BibitemShut {NoStop}%
\bibitem [{\citenamefont {Kamsma}\ \emph {et~al.}(2025)\citenamefont {Kamsma}, \citenamefont {Klop}, \citenamefont {Boon}, \citenamefont {Spitoni}, \citenamefont {Rueckauer},\ and\ \citenamefont {Van~Roij}}]{Kamsma2025ChemicallyApplications}%
  \BibitemOpen
  \bibfield  {author} {\bibinfo {author} {\bibfnamefont {T.~M.}\ \bibnamefont {Kamsma}}, \bibinfo {author} {\bibfnamefont {M.~S.}\ \bibnamefont {Klop}}, \bibinfo {author} {\bibfnamefont {W.~Q.}\ \bibnamefont {Boon}}, \bibinfo {author} {\bibfnamefont {C.}~\bibnamefont {Spitoni}}, \bibinfo {author} {\bibfnamefont {B.}~\bibnamefont {Rueckauer}},\ and\ \bibinfo {author} {\bibfnamefont {R.}~\bibnamefont {Van~Roij}},\ }\bibfield  {title} {\bibinfo {title} {{Chemically regulated conical channel synapse for neuromorphic and sensing applications}},\ }\href {https://doi.org/10.1103/PHYSREVRESEARCH.7.013328/SM.PDF} {\bibfield  {journal} {\bibinfo  {journal} {Physical Review Research}\ }\textbf {\bibinfo {volume} {7}},\ \bibinfo {pages} {013328} (\bibinfo {year} {2025})}\BibitemShut {NoStop}%
\bibitem [{\citenamefont {Zahedi}\ \emph {et~al.}(2020)\citenamefont {Zahedi}, \citenamefont {Mayahinia}, \citenamefont {Lebdeh}, \citenamefont {Wong},\ and\ \citenamefont {Hamdioui}}]{Zahedi2020EfficientCim}%
  \BibitemOpen
  \bibfield  {author} {\bibinfo {author} {\bibfnamefont {M.}~\bibnamefont {Zahedi}}, \bibinfo {author} {\bibfnamefont {M.}~\bibnamefont {Mayahinia}}, \bibinfo {author} {\bibfnamefont {M.~A.}\ \bibnamefont {Lebdeh}}, \bibinfo {author} {\bibfnamefont {S.}~\bibnamefont {Wong}},\ and\ \bibinfo {author} {\bibfnamefont {S.}~\bibnamefont {Hamdioui}},\ }\bibfield  {title} {\bibinfo {title} {{Efficient organization of digital periphery to support integer datatype for memristor-based cim}},\ }\href {https://doi.org/10.1109/ISVLSI49217.2020.00047} {\bibfield  {journal} {\bibinfo  {journal} {Proceedings of IEEE Computer Society Annual Symposium on VLSI, ISVLSI}\ }\textbf {\bibinfo {volume} {2020-July}},\ \bibinfo {pages} {216} (\bibinfo {year} {2020})}\BibitemShut {NoStop}%
\bibitem [{\citenamefont {Fraiman}\ \emph {et~al.}()\citenamefont {Fraiman}, \citenamefont {Sabbagh}, \citenamefont {Yossifon},\ and\ \citenamefont {Fish}}]{FraimanTowardComponents}%
  \BibitemOpen
  \bibfield  {author} {\bibinfo {author} {\bibfnamefont {N.~E.}\ \bibnamefont {Fraiman}}, \bibinfo {author} {\bibfnamefont {B.}~\bibnamefont {Sabbagh}}, \bibinfo {author} {\bibfnamefont {G.}~\bibnamefont {Yossifon}},\ and\ \bibinfo {author} {\bibfnamefont {A.}~\bibnamefont {Fish}},\ }\bibfield  {title} {\bibinfo {title} {{Toward an Ion-Based Large-Scale Integrated Circuit: Circuit Level Design, Simulation, and Integration of Iontronic Components}},\ }\href@noop {} {\bibfield  {journal} {\bibinfo  {journal} {arXiv}\ }\textbf {\bibinfo {volume} {2412.07784}}}\BibitemShut {NoStop}%
\end{thebibliography}

%

\end{document}